\documentclass[fleqn, 1p, sort&compress]{elsarticle}

\usepackage{hyperref}
\graphicspath{ {images/} }
\usepackage{mathtools}
\usepackage{adjustbox}
\usepackage{booktabs}
\usepackage{diagbox}
\usepackage{floatpag}
\usepackage{tikz}
\usepackage{graphicx, subcaption}
\usepackage{multirow}
\usepackage{amsmath, bm}
\usepackage{array}
\usepackage{booktabs}
\usepackage{setspace}
\usepackage[T1]{fontenc}
\usepackage{changepage}
\usepackage{fancyhdr}

\usetikzlibrary{shapes.geometric, arrows, calc}
\tikzstyle{block} = [rectangle, rounded corners, minimum width=1cm, minimum height=2.25cm,text centered, draw=black]
\tikzstyle{arrow} = [thick,->,>=stealth]

\DeclarePairedDelimiter\floor{\lfloor}{\rfloor}
\journal{Speech Communication}

\begin{document}

\begin{frontmatter}

\title{Modulation spectral features for speech emotion recognition using deep neural networks}

\author[mysecondaryaddress]{Premjeet Singh\corref{corr1}}
\ead{premsingh@iitkgp.ac.in}
\address[mysecondaryaddress]{Department of Electronics \& Electrical Communication Engineering \\ Indian Institute of Technology, India-721302, Kharagpur, India }

\author[mymainaddress]{Md Sahidullah}
\ead{md.sahidullah@inria.fr}

\author[mysecondaryaddress]{Goutam Saha}
\ead{gsaha@ece.iitkgp.ernet.in}

\address[mymainaddress]{Universit\'{e} de Lorraine, CNRS, Inria, LORIA, F-54000, Nancy, France}

\cortext[corr1]{Corresponding author}

\begin{abstract}
This work explores the use of constant-Q transform based modulation spectral features (CQT-MSF) for speech emotion recognition (SER). The human perception and analysis of sound comprise of two important cognitive parts: early auditory analysis and cortex-based processing. The early auditory analysis considers spectrogram-based representation whereas cortex-based analysis includes extraction of temporal modulations from the spectrogram. This temporal modulation representation of spectrogram is called modulation spectral feature (MSF). As the constant-Q transform (CQT) provides higher resolution at emotion salient low-frequency regions of speech, we find that CQT-based spectrogram, together with its temporal modulations, provides a representation enriched with emotion-specific information. We argue that CQT-MSF when used with a 2-dimensional convolutional network can provide a time-shift invariant and deformation insensitive representation for SER. Our results show that CQT-MSF outperforms standard mel-scale based spectrogram and its modulation features on two popular SER databases, Berlin EmoDB and RAVDESS. We also show that our proposed feature outperforms the shift and deformation invariant scattering transform coefficients, hence, showing the importance of joint hand-crafted and self-learned feature extraction instead of reliance on complete hand-crafted features. Finally, we perform Grad-CAM analysis to visually inspect the contribution of constant-Q modulation features over SER.
\end{abstract}


\end{frontmatter}

\thispagestyle{fancy}
\fancyhf{}
\chead{\footnotesize Accepted for publication in Speech Communication Journal}
\lfoot{\footnotesize \copyright~2022. This manuscript version is made available under the CC-BY-NC-ND 4.0 license.
\url{https://creativecommons.org/licenses/by-nc-nd/4.0/}}
\renewcommand{\headrulewidth}{0pt}


\section{Introduction}
\label{sec1}
Speech emotion recognition (SER) is the process of automatic prediction of speaker's emotional state from his/her speech samples. A speech sample generally remains enriched with various information, such as speaker, language, emotion, context, recording environment, gender and age, intricately entangled to each other~\cite{krothapalli2013speech}. Human mind is congenitally trained to disentangle such information, however, same is not true for machines~\cite{picard2003affective}. Machines need to be specifically trained to extract cues pertaining to a particular information. Among such, extraction of emotion-specific cues for SER is still considered a challenging task. The challenge basically persists because of the differences in the manner of emotion expression across individuals \cite{el2011survey}. These differences stem from factors such as speaker's culture and background, ethnicity, speaker's mood, gender, manner of speech, etc.~\cite{el2011survey, SHAHFAHAD2021}. For automatic SER, a machine should be capable of extracting emotion-specific cues in the presence of all such variabilities. \\

SER finds application in several human-computer interaction domains such as sentiment analysis in customer service, health care systems, self-driving vehicles, auto-pilot systems, product advertisement and analysis~\cite{krothapalli2013speech, el2011survey, akccay2020speech}. One of the first seminal works in SER was aimed towards emotion information extraction using different speech cues~\cite{dellaert1996recognizing}. Various works that followed discovered that \emph{speech prosody} (pitch, intonation, energy, loudness, etc.) contain significant information for emotion discrimination~\cite{eyben2010towards,eyben2015geneva,chen2012speech}. Similarly, several other works report that \emph{spectral features} (spectral flux, centroid, mel-frequency cepstral coefficients (MFCCs), etc.) and \emph{voice-quality features} (jitter, shimmer, harmonic-to-noise ratio (HNR), etc.) of speech are also important for SER~\cite{li2007stress}. For classification, these extracted features are processed with a classifier back-end such as \emph{support vector machine} (SVM), \emph{Gaussian mixture model} (GMM), and \emph{k-nearest neighbour} (k-NN) for emotion class prediction. These approaches which employ certain signal processing algorithm for feature extraction are termed hand-crafted feature based approaches for SER. Hand-crafted approaches enjoy the advantage of being interpretable, in terms of which feature or speech characteristic is more relevant for emotions, and are computationally inexpensive. However, hand-crafted features often suffer from \emph{curse of dimensionality}, especially when \emph{brute-force} method based SER system is used~\cite{batliner2011whodunnit}. \\

Recent advancements in signal processing have introduced \emph{deep neural networks} (DNN) into the speech processing domain. DNNs have the impressive ability by which, given the required data, they automatically learn to obtain a possible solution to pattern recognition problems. This is accomplished by automatically updating the DNN parameters so as to reduce the defined loss function and approach towards the local minima. 
In SER system, deep networks are either used as automatic feature extractors or as classifiers for emotion class prediction.
Recently, a new deep learning paradigm is also introduced which performs both feature extraction and emotion classification in an end-to-end fashion. Along these lines, several works use \emph{convolutional neural network} (CNN) as automatic feature extractor for SER~\cite{zhang2017speech, mao2014learning}. In contrast, other approaches use hand-crafted methods for feature extraction which are then used as features for DNN classifier input~\cite{ghosh2016representation, zhao2019speech, issa2020speech}. To obtain an end-to-end solution for SER works in~\cite{trigeorgis,tzirakis2018end,tang2018end} have used DNNs where the initial layers extract the emotion-relevant features and final layers act as classifier. In recent years, deep learning methods have been consistently shown to outperform hand-crafted feature based SER techniques. \\

In spite of their tremendous success, DNNs have major practical disadvantages. One such disadvantage is the requirement of large labelled database for proper DNN training~\cite{rolnick2017deep}. In contrast to other speech classification problems, such as speech and speaker recognition, large speech corpora are not available for evaluating SER task. Various ethical and legal issues make it difficult to collect large dataset of natural emotional voices from real-world scenario~\cite{el2011survey, akccay2020speech}. To somewhat alleviate this issue, acted emotion recordings are generally used where skilled actors enact a predefined set of emotions. However, this approach is not considered very appropriate as acted emotions are often exaggerated versions of natural emotions~\cite{el2011survey, akccay2020speech}. Another disadvantage of DNN is its complexity. Due to large trainable parameter set and high non-linear relationship between input and output, DNNs are often termed \emph{black-box} models, which are very difficult to understand/interpret~\cite{lipton2018mythos, BARREDOARRIETA202082, kimura2020new}. As the training includes optimization of all DNN parameters, it takes much larger time for DNNs to train as compared to classical statistical methods (e.g., SVM or GMM). Hence, even though very appealing, DNN models are still far-off a completely optimized SER approach. \\

The above discussion leads to the conclusion that both hand-crafted and DNN based feature extraction methods have their own set of advantages and disadvantages. In this work, we aim to exploit the advantages of both the methods for improved SER performance. Our framework is similar to the combination of hand-crafted feature, in the form of time-frequency representation, and a DNN model for further feature enrichment as used in other SER works~\cite{ghosh2016representation, zhao2019speech, issa2020speech}. However, our approach incorporates the prior (domain) knowledge of speech processing in humans, i.e., early auditory and cortex-based processing of speech \cite{sshamma}, for an improved hand-crafted feature representation. Being data-driven, DNN-based machine learning approaches suffer in performance, especially when there are constraints over the training data, e.g., limited size, ethical concerns in recording and poor quality of data \cite{muralidhar2018incorporating}, all of which are relevant for SER databases. Evidences reveal that such disadvantages can be alleviated by the use of domain knowledge \cite{muralidhar2018incorporating, rueden2021} in hand-crafted feature generation. Further, regarding speech representations, spectrogram and mel-spectrogram are considered the \emph{de-facto} standard of time-frequency representations in SER. However, they encompass only the early auditory processing of speech and lack cortical information. \\

\begin{figure}[t!]
    \centering
    \includegraphics[scale=0.65]{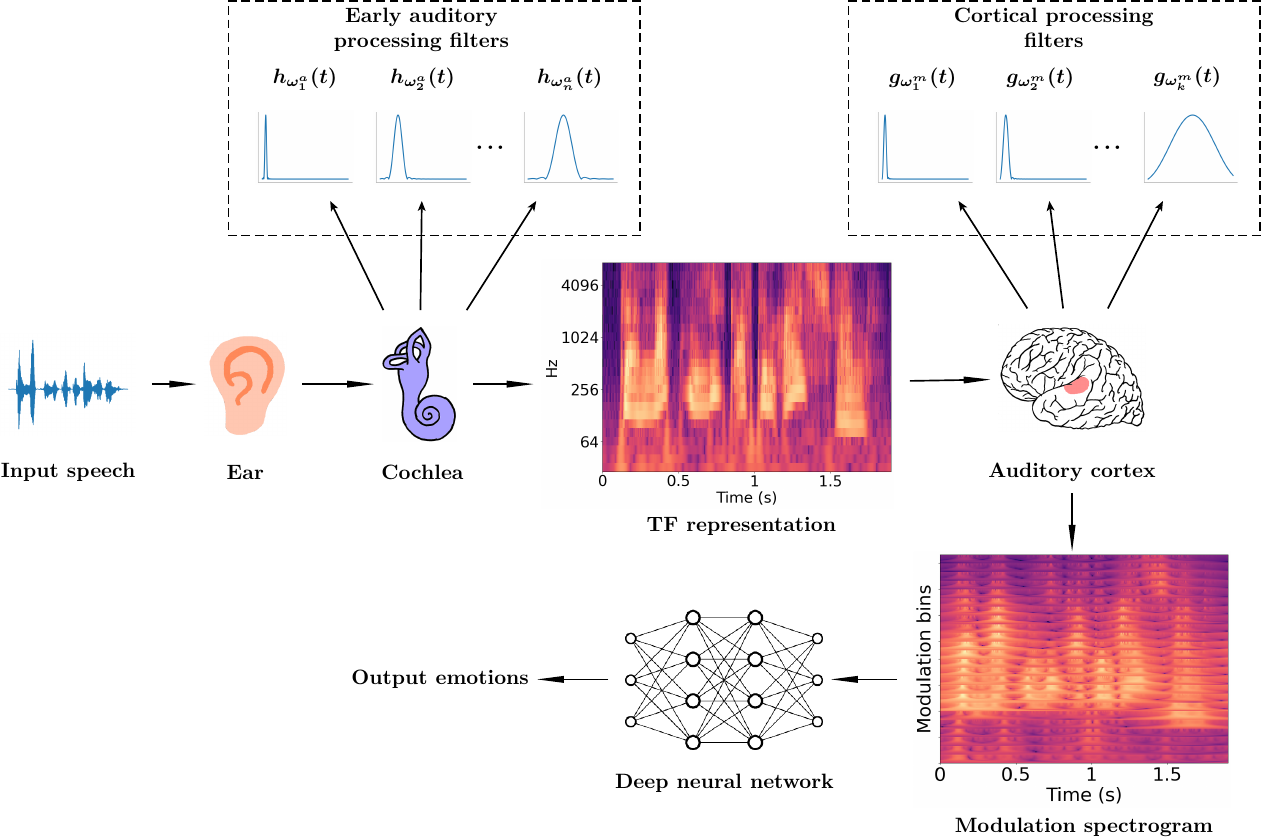}
    \caption{The two-staged speech processing in the human auditory system for SER. The input speech captured at ear is converted to a form similar to time-frequency (TF) representation by the early auditory processing filters present in cochlea. This representation is then passed on to the auditory cortex in the brain for processing with cortical filters. The highlighted part in brain image identifies the auditory cortex region of the brain. The cortical filter processing leads to a modulation spectrogram based representation. This is further processed by the inner regions of the brain to finally decode emotions. Our employed deep neural network, which is loosely based on the studies of the brain and nervous system, models the inner processing of the brain to identify the emotion classes from the input modulation spectrogram feature. The $h_{\omega_n^a}(t)$ and $g_{\omega_k^m}(t)$ depict the impulse response of $n$th early auditory and $m$th cortical processing filter, respectively. The Figure shows logarithm applied TF and modulation spectrogram representation.}
    \label{flow_diag}
\end{figure}

Inspired by this fact, we first employ a hand-crafted feature extraction technique which combines an emotion relevant early auditory representation with corresponding cortex-based representation of the speech for SER. These features are then processed by a deep convolutional neural network which further extracts the emotion relevant information. Two machine learning frameworks are used at the back-end: Convolutional network with fully connected layer, and convolutional layer for embedding extraction with SVM classifier for final emotion class prediction. Such combination of multi-stage hand-crafted feature with DNN at back-end more closely follows the natural speech processing workflow in humans where the auditory system captures the signal and extracts the features, which are then transmitted to the inner regions of brain for further analysis and understanding. The achieved improvement in performance over different databases further consolidates our hypothesis of two-staged hand-crafted speech processing for SER. Figure~\ref{flow_diag} provides a general overview of the two-staged processing framework in human auditory system for SER.  \\

In the next section (Section~\ref{lit_review}), we describe the relevant literature and discuss the motivation and major contributions of this work. Section~\ref{cqt_section} and \ref{mod_spec_section} provides a brief introduction to the early auditory and cortex-based feature representations used in this work. Section~\ref{exp_setup} describes the experimental setup used to perform the experiments. Section~\ref{res_and_dis} describes the results obtained with the proposed feature and comparison with the standard features followed by corresponding discussion. Finally, Section~\ref{conc_sec} includes the conclusive statements of the work.

\section{Related Works and Motivation}
\label{lit_review}

In this section, we provide a brief review of works related to the frequency localisation of emotions. We then discuss some works which describe the relevance of modulation spectrogram in speech processing. This is followed by description of the motivation of our proposed feature and the major contributions of this work.

\subsection{Literature Review}

Several studies, aimed towards analysing the importance of spectral frequencies, have reported the prominence of low frequencies in SER. Authors in~\cite{bou2000comparative} report the prominence of first formant frequency (F1) for recognition of \emph{Anger} and second formant (F2) for recognition of \emph{Neutral}. Studies performed in~\cite{goudbeek2009emotion} found that high arousal emotions, e.g, \emph{Anger}, \emph{Happy}, have higher average F$1$ value and lower F$2$ value. They also found that positive valence emotions (e.g., \emph{Happy}, \emph{Pride}, \emph{Relief}) have higher average F$2$ value. Authors in~\cite{bozkurt2011formant} also report discrimination between idle and negative emotions using the temporal patterns of first two formant frequencies. In~\cite{lech2018amplitude}, authors show that non-linear frequency scales (e.g., equivalent rectangular bandwidth (ERB), mel, logarithmic) when applied for sub-band partitioning and energy computation over discrete Fourier transform based spectrogram, results in improved SER accuracy. Such studies hint toward the requirement of a non-linear frequency scale based time-frequency representation with higher emphasis on low-frequency regions of speech. \\

Regarding human sound perception, evidences in literature suggest that the process of auditory signal analysis can be modelled into two stages: (i)~\emph{Early auditory stage}, which models the incoming audio signal into a spectrogram based representation. (ii)~\emph{Cortical analysis stage}, which extracts the spectro-temporal modulation relationship among different audio cues from the auditory spectrogram \cite{sshamma,tchi}. Such modelling strategy has been found effective in the analysis of both speech and music signals~\cite{sshamma}. The spectral and temporal modulation features of speech spectrogram are also highly related to speech intelligibility, noise and reverberation effects~\cite{tchi}. In~\cite{kumar2012features}, authors report that the spectro-temporal representation of audio (non-speech) signals with positive\textbackslash negative valence is different from that of neutral sounds. They also report that spectral frequency and temporal modulation frequency can represent the valence information of sounds. 

Authors in~\cite{ARNAL20152051} compared the temporal modulations of human speech with scream voice and concluded that slow temporal variations ($< 20$~Hz) contain most linguistic (both prosodic and syllabic cues) information. \\

In speech analysis, temporal modulation features are called \emph{modulation spectral features} (MSFs). Owing to their relatedness to speech intelligibility, MSFs have been extensively used in speech processing. Some works also successfully explored modulation features for speaker identification, verification and audio coding~\cite{vuuren1998importance, nsephus}. Author in~\cite{hermanskyhistory} provides a comprehensive description of the history of the use of modulation features in speech recognition. \\

Modulation features have also been explored for emotion recognition in speech. Authors in~\cite{wu2011automatic} used MSF for emotion recognition and provided a detailed explanation of its relevance for SER. In~\cite{alam2013amplitude}, authors used a smoothed nonlinear operator to obtain the amplitude modulated power spectrum of the gammatone filterbank generated spectrogram and showed improvement over standard MFCC for SER. Authors in~\cite{Zhu+2016} studied the relationship between human emotion perception and the MSFs of emotional speech and concluded on the suitability of modulation features for emotion recognition. Authors in~\cite{peng2020} used 3-D convolutions and attention-based recurrent networks to combine auditory analysis and attention mechanisms for SER. This work also explains that the temporal modulations extracted from auditory analysis contain periodicity information important for emotion recognition. In~\cite{avila}, various feature pooling, such as \emph{mean}, \emph{standard deviation}, and \emph{kurtosis}, on frame-level measure of MSF to be used for ``in-the-wild'' dimension-based SER (dimensional SER includes projection of speech onto three emotion dimensions: \emph{valence}, \emph{arousal} and \emph{dominance)}. The authors report improvement in results over frame-wise modulation spectral feature baseline for various noise and reverberated speech scenarios. Similar MSF measures when used with Bag-of-Audio-Words (BoAW) approach showed SER improvement against environmental noise in~\cite{kshirsagar2022}. In~\cite{avila2019}, authors use modulation spectral features with convolutional neural networks to discriminate between stress-based speech and neutral speech. The authors show that the modulation spectral features when used with CNN with the time frames intact (without statistics pooling over time frames of MSF) gives better performance, especially over increased number of target emotion classes. In~\cite{yeh2010spectro}, authors show that joint spectro-temporal modulation representation outperforms standard MFCC in emotion classification of noisy speech. Recently, the authors in~\cite{PENG2021261} have also used MSF over cochleagram features with a long short term memory (LSTM) based system for dimensional SER. The work explains that arousal information can be characterised by the amplitude envelope of speech signal, whereas valence information is characterised by the temporal dynamics of amplitude envelope. Since it is difficult to obtain such dynamics from low-level descriptor (LLD) features, auditory analysis based temporal modulation features can potentially represent the required temporal dynamics for SER.

\subsection{Motivation and Contributions}

The literature in SER reveals two important speech characteristics for emotion prediction: the importance of low frequencies, and the importance of temporal modulations of spectrogram. To address the importance of low-frequency information, we use constant-Q transform (CQT) based time-frequency representation for SER. CQT provides higher frequency resolution and increased time invariance at low frequencies thereby emphasizing the low-frequency regions of speech~\cite{SINGH2022103712}. This helps in better resolution of emotion salient frequency regions of speech and improved SER performance~\cite{singh2021}. CQT is also known to provide a representation with visible pitch frequency and well-separated pitch harmonics~\cite{chandra}. Because of high relevance of pitch information in emotion discrimination, this property of CQT makes it more suitable for SER over standard mel-based features. \\

To further enhance the CQT-based system while utilising the understanding of domain knowledge of human auditory-cortical physiology~\cite{tchi}, we propose to use temporal modulation of CQT spectrogram representation for SER. Specifically, we use CQT spectrogram representation for auditory analysis and extract temporal modulations of CQT by again using constant-Q filters, for cortical analysis. In this way, we obtain the temporal modulation of emotion salient low-frequency regions which are emphasized by CQT. Studies show that such use of constant-Q modulation filterbank better approximates the cortical sound processing in humans~\cite{suki2002, suki2004}. The constant-Q factor characteristic of modulation filters also lead to higher resolution at lower modulation frequencies, hence, providing an arrangement that helps in identifying any deviation from general (or \emph{Neutral}) speech modulation rate ($2$-$4$~Hz)~\cite{hermansky2011speech}. Our choice of constant-Q filters in both stages is also inspired from the study of early auditory and cortical stages of mammalian auditory cortex~\cite{tchi, zotkin2003}. We term our proposed feature as \emph{constant-Q transform based modulation spectral feature} (CQT-MSF). A 2-dimensional convolution neural network architecture (2-D CNN) is used to further refine the emotion information present in CQT-MSF feature. We compare the performance of CQT-MSF with mel-frequency spectral coefficients (MFSC) and show that the constant-Q non-linearity based auditory-cortical features outperform the mel-scale non-linearity based features. We also investigate the performance differences obtained with auditory and cortical representations taken separately. We also highlight the striking similarity of CQT-MSF with the wavelet-based time-shift and deformation invariant coefficients, known as scattering transform coefficients~\cite{anden}. Our main contributions in this work are as follows:

\begin{itemize}

    \item This study proposes a new human auditory-cortical physiology based SER framework.

    \item We propose a modulation feature extraction technique using constant-Q filterbank over constant-Q spectrogram and analyse its relevance from vocal emotion perspective.

    \item We perform similarity analysis with another two-staged auditory-cortical feature representation: Scattering transform.

    \item We also perform explainability analysis to visually inspect different regions of CQT-MSF that weigh the most in prediction of a particular emotion class.

    \item The study further hints correlation between music training and emotion understanding by discussing the case of \emph{Amusia}~\cite{sydney2021}, and the possible analogy between modulation computed over CQT spectrogram and the cortex-level processing of sound in music trained individuals~\cite{dmitrieva2006ontogenetic, fuller2014musician, twaite2016examining, weijkamp2017attention, thompson2004decoding, nussbaum2021}.

\end{itemize}

\section{Early auditory processing: Constant-Q Transform (CQT)}
\label{cqt_section}

Our proposed features are based on the use of constant-Q filterbanks for both time-frequency (early auditory) and temporal modulation (cortical) based analysis of speech. In this section, we briefly discuss the CQT method of time-frequency representation. CQT uses constant \emph{quality factor} (Q-factor) bandpass filters with logarithmically spaced center frequencies~\cite{todisco2017constant}. Mathematical formulation of constant-Q transform is given by,

\begin{equation}
    X^{CQT}[k,n]~=\sum_{j~=~n-\floor{N_K/2}}^{n+\floor{N_k/2}}~x(j)a_k^*(j-n+N_k/2)
    \label{eq:CQT}
\end{equation}

\noindent where $k$ denotes the CQT frequency index, $\floor{.}$ denotes the rounding-off to nearest integer towards negative infinity and $a_k^*(n)$ is the complex conjugate of the CQT basis function for $k$\textsuperscript{th} CQT bin. The CQT basis, or the time-frequency \emph{atom}, is a complex time domain waveform given as,

\begin{equation}
    a_k(n)~=~ \frac{1}{N_k}w\left(\frac{n}{N_k}\right) exp\left[-i2\pi n\frac{f_k}{f_s}\right]
\end{equation}

\noindent where $f_k$ is the center frequency of $a_k$, $f_s$ is the sampling frequency and $w(n)$ is the window function with length $N_k$. We use the standard \emph{Hann} window in this work for CQT computation. The center frequencies of filters in constant-Q transform are spaced by the relation $f_k~=~f_{\mathrm{min}}2^\frac{k-1}{B}$ where $f_k$ is the frequency of $k$th filterbank, $f_{\mathrm{min}}$ being the frequency of the lowest bin and $B$ the number of frequency bins used per octave of frequency. This binary logarithmic spacing leads to more frequency bins at lower frequencies, as compared to high frequencies, and hence provides higher frequency resolution at low frequencies~\cite{todisco2017constant}. In time domain, such filters can be given as truncated sinusoids (e.g., truncated with \emph{Hann} window) with different lengths~\cite{schorkhuber2010constant}, given by,

\begin{equation}
N_k~=~\frac{q f_s}{f_k(2^{\frac{1}{B}}-1)} 
\label{eq:scale}
\end{equation}

\renewcommand{\thesubfigure}{\alph{subfigure}}
\begin{figure}[t]
\centering
\begin{subfigure}[t]{0.5\textwidth}
    \centering
    \hbox{\hspace{-0.7cm}\includegraphics[scale=0.23]{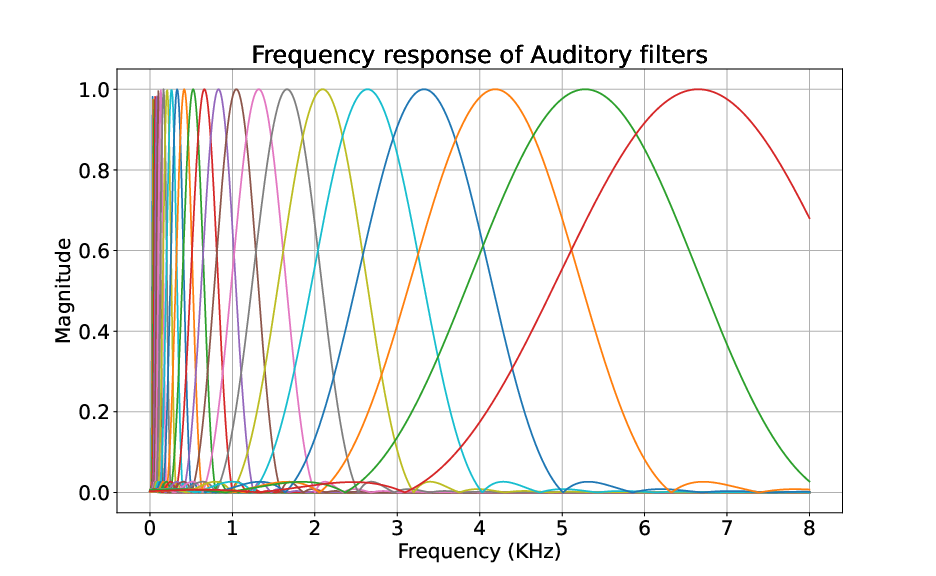}}
    \caption{Early auditory filterbank}
    \label{early_aud_filters}
\end{subfigure}%
\begin{subfigure}[t]{0.5\textwidth}
    \centering
    \includegraphics[scale=0.23]{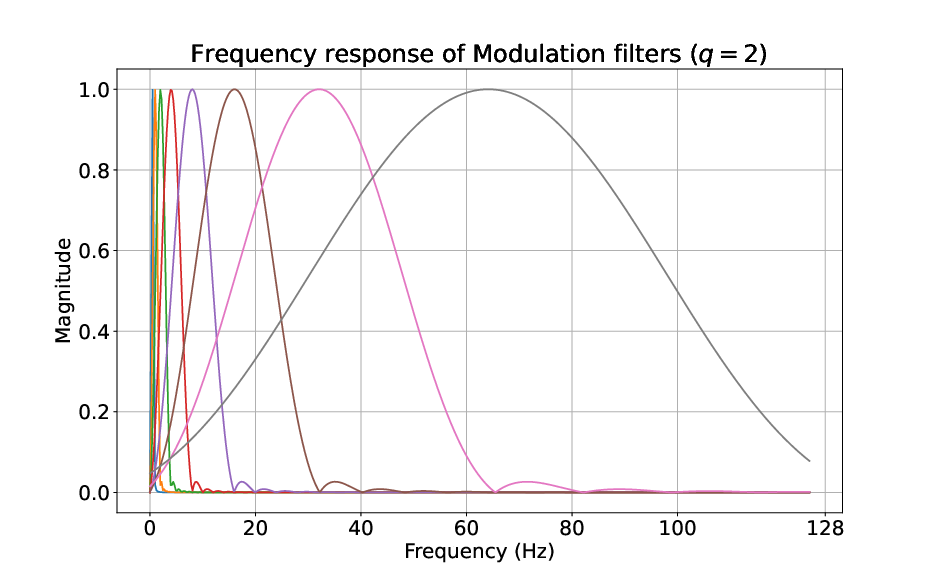}
    \caption{Modulation filterbank}
    \label{cortical_filters}
\end{subfigure}%
\caption{Auditory and Modulation filter banks used in CQT-MSF. The modulation filters shown here have scale factor ($q$) value 2.}
\label{CQT-MSF filterbanks}
\end{figure}

\noindent where $q$ is the filter scaling factor. This scaling factor offers to change the time (and hence frequency) resolution of CQT bases without affecting $B$~\cite{schorkhuber2010constant}. When compared with mel-based features, the mel-scale is also logarithmic in nature. However, mel-scale uses a decadic logarithm scale (or natural logarithm in some implementations), because of which, the emphasis on low-frequencies is not as prominent as CQT.  \\

The computation of CQT, as described in Eq.~\ref{eq:CQT}, includes convolution of \emph{atom} with every time sample of the input signal. However, the fast CQT computation algorithm~\cite{schorkhuber2010constant} introduced a \emph{hop length} parameter, which describes the number of samples the time window is shifted for next time frame CQT computation. The hop length is kept equal to integer multiples of $2^ {\mathrm{No. \  of \ octaves}}$ so that the corresponding signal frames at different frequencies do not fall out of alignment \cite{schorkhuber2010constant}. In CQT representation, the number of octaves is given by $\log_{2}{\frac{F_{\mathrm{max}}}{F_{\mathrm{min}}}}$ \cite{todisco2017constant} where $F_{\mathrm{min}}$ and $F_{\mathrm{max}}$ are the minimum and maximum frequency of operation, respectively. For CQT computation in this work, we use the \emph{LibROSA}\footnote{\url{https://librosa.github.io/}} toolkit~\cite{brian_mcfee_2021_4792298} which follows all the computational details of the fast CQT implementation mentioned above.

\section{Cortex-based processing: Modulation Spectrogram}
\label{mod_spec_section}

Modulation spectrogram shows the temporal variation pattern of the spectral components in spectrogram. According to~\cite{hermansky2011speech}, speech signal is composed of two parts, the carrier, i.e., the vocal cords excitation, and the varying modulation envelope which is the result of changes in orientation of different vocal organs over time. The low-frequencies of the modulation envelope characterise slow variations of the complete spectral structure, which is known to encode most of the phonetic information~\cite{greenberg1997, hermanskyhistory, elhilali2019modulation}. Let $S(t, ~\omega)$ be the speech spectrogram. The temporal evolution of a frequency bin $\omega_o$ in $S(t, ~\omega)$, over time $t$, is a one-dimensional time-series. The spectral representation of this time-series $S(t, ~\omega_o)$ constitutes the modulation spectrum of frequency bin $\omega_o$ over $T$, where $T$ is the spectrogram time window (with duration equal to window length $N_k$). \\

For speech, most of the modulation energy remains concentrated around $2$-$4$~Hz range with peak at $4$~Hz~\cite{hermansky2011speech}. This makes $4$~Hz to be considered as the syllabic rate of normal (\emph{Neutral}) speech. Deviations from this rate generally result from infliction of noise or reverberation effects over speech~\cite{elhilali2019modulation, moritz2011}. It is studied in SER literature that rate of speech is higher than \emph{Neutral} class for high arousal emotions, such as, \emph{Anger}, \emph{Fear} and lower for low arousal emotions, such as \emph{Sad}, \emph{Boredom}~\cite{banse1996acoustic}. 
Hence, this deviation of modulation energy peak from $4$~Hz can be used for emotion discrimination over arousal scale.

\begin{figure}[t!]
    \centering
    \includegraphics[scale=0.66]{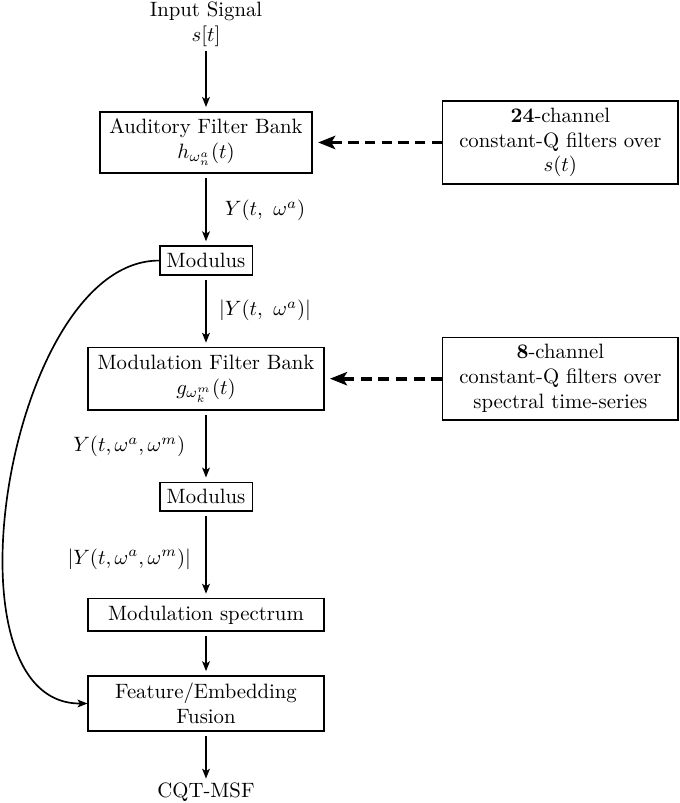}
    \caption{Block diagram of proposed CQT-MSF feature extraction method. In figure, $\omega^a$ refers to acoustic frequency and $\omega^m$ refers to the modulation frequency.}
    \label{block_diagram}
\end{figure}

\subsection{Constant-Q based Modulation Spectral Features (CQT-MSF)}
\label{cqmsf}

\begin{figure}[t]
    \centering
    \hbox{\hspace{-0.4cm}\includegraphics[scale = 0.35]{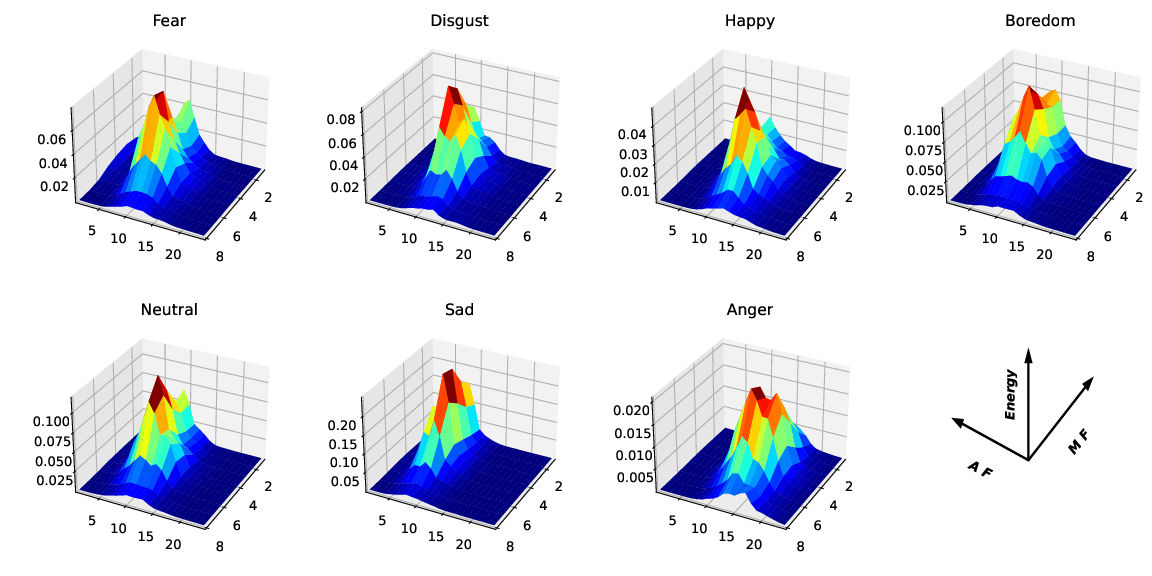}}
    \caption{Modulation spectral features (averaged over time) for different emotions in EmoDB. The `\textbf{M F}' refers to the modulation frequency channels and `\textbf{A F}' refers to the acoustic frequency channels. The modulation filters used in this analysis has filter scale ($q$) value 2.}
    \label{cqt_msf_analysis}
\end{figure}


In this subsection, we compute modulations of CQT bins and combine them with CQT spectrogram to generate CQT-MSF. The first stage early auditory analysis (CQT spectrogram) in the CQT-MSF feature can be given as,
 
\begin{equation}
     Y(t, ~\omega^a) = s(t) * h_{\omega_n^a}(t); \  \omega_0^a \leq \omega_n^a < \omega_C^a,
\end{equation}

\noindent where, $s(t)$ is the input speech signal, $h_{\omega_n^a}(t)$ is the impulse response of $n$th constant quality factor auditory filter with $\omega_n^a$ center frequency, $C$ is the number of auditory filters and $Y(t, ~\omega^a)$ is the corresponding time-frequency representation. Fig.~\ref{early_aud_filters} shows the frequency response of different $h_{\omega_n^a}(t)$ used. For envelope extraction, modulus operation is applied over $Y(t, ~\omega^a)$, i.e., $|Y(t, ~\omega^a)|$. The resulting representation provides the temporal trajectories of different frequency bins in $Y(t, ~\omega^a)$. \\

For cortical analysis, the $|Y(t, ~\omega^a)|$ is passed through a modulation filterbank. The modulation spectrogram computed over time-frequency representation $|Y(t, ~\omega^a)|$, is given as,

\begin{equation}
    Y(t, \omega^a, \omega^m) = |Y(t, ~\omega_n^a)| * g_{\omega_k^m}(t); \ \omega_0^m \leq \omega_k^m < \omega_M^m, ~\omega_0^a \leq \omega_n^a <\omega_C^a,
    \label{eq2}
\end{equation}

\noindent where, $g_{\omega_k^m}(t)$ is the impulse response of $k$th modulation filter with $\omega_k^m$ center frequency and $M$ is the total number of modulation filters. Similar to the output of the first stage, we use the modulus of computed modulation spectrum coefficients computed over all frequency bins of CQT spectrogram, i.e., $|Y(t, \omega^a, \omega^m)|$~\cite{hermansky2011speech}. Fig.~\ref{block_diagram} shows the block diagram of CQT-MSF feature extraction. The complete MSF includes concatenation of temporal modulations, computed using every modulation filter ($g_{\omega_0^m} \leq g_{\omega_k^m} < g_{\omega_M^m}$), of all frequency bins in the time-frequency representation, i.e., for $\omega_0^a \leq \omega_n^a < \omega_C^a$ bins where $C = 24$ auditory channels in our experiments. Regarding the properties of MSF, study performed in~\cite{tchi} report distinction between three different temporal modulation rates: slow, intermediate, and fast. The slow modulation rate is shown to roughly correspond to the syllable or speaking rate. Whereas the intermediate modulation rate appearing because of interharmonic interaction is shown to reflect the fundamental frequency of the signal. This shows the importance of temporal modulation for pitch representation, and hence, SER. Temporal modulation extracted by MSF represent tempo~\cite{DING2017181}, pitch, and timber~\cite{zotkin2003}, all of which are related to emotion information in speech. \\

Fig.~\ref{cqt_msf_analysis} shows the time-averaged CQT-MSF coefficients for utterances of different emotion classes of the EmoDB database. The MF and AF refer to the modulation and auditory frequency channels, respectively. In terms of modulation frequency, the highest peak in \emph{Neutral} emotion is observed at $4$~Hz modulation frequency with another peak around $0.5$~Hz. Compared to \emph{Neutral} class, low arousal emotions (\emph{Boredom} and \emph{Sad}) also have energies extending towards $0$-$4$~Hz modulation frequency range. High arousal emotions (\emph{Anger}, \emph{Fear}) have peak around $4$-$8$~Hz modulation frequency range. In contrast, \emph{Happy} also has a peak at $4$~Hz similar to \emph{Neutral}. Similarly, \emph{Disgust} also shows a peak at $4$~Hz followed by another peak at $2$~Hz. From AF perspective, \emph{Anger} emotion shows peak at high frequencies (high AF), whereas, in \emph{Sad} low auditory frequencies are more dominant. For remaining emotions, auditory energy distribution extends almost similarly over mid-auditory frequencies. This analysis shows the higher emotion discrimination potential of combined MF and AF channels as compared to only AF channel-based representation. \\

\begin{figure}[t!]
    \centering
    \hbox{\hspace{-1cm}{\includegraphics[scale = 0.4]{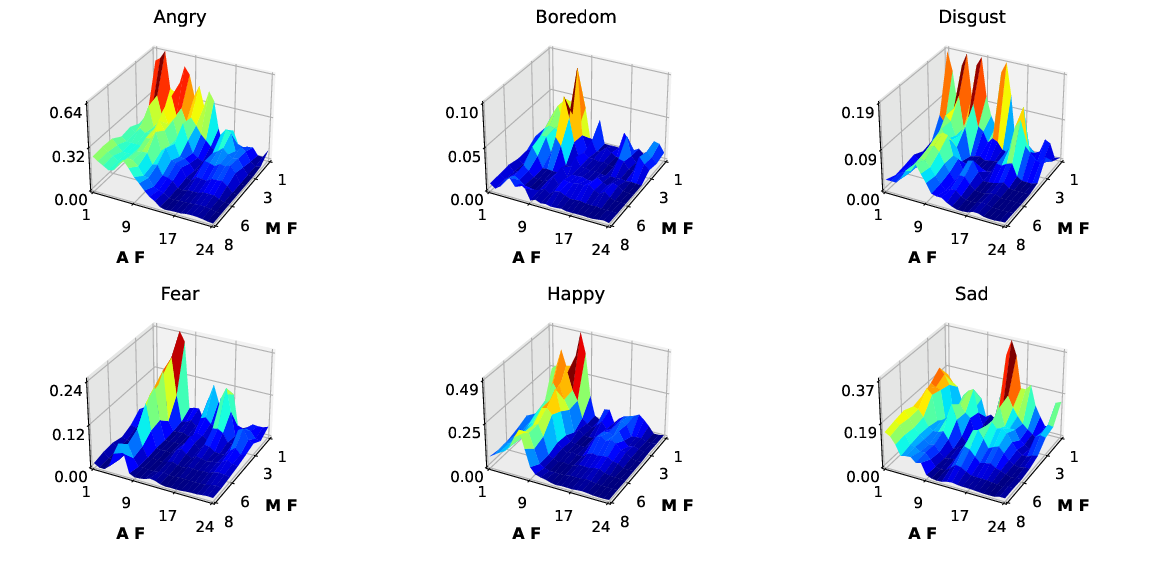}}}
    \caption{F-ratio values of different auditory frequency (AF) and modulation frequency (MF) bins of modulation spectrum features computed over EmoDB database. The F-ratio is calculated over time-averaged modulation spectrum features (MSF) between \emph{Neutral} and every other emotion class. The modulation filters used in this analysis have filter scale ($q$) value 2.}
    \label{f_ratio_modspec}
\end{figure}

To further analyze the discriminative potential of modulation spectrum features, we perform F-ratio analysis between the time-averaged modulation features of various emotion classes and the \emph{Neutral} class of the EmoDB database~\cite{LU2008, dipjyoti2017}. Fig.~\ref{f_ratio_modspec} shows the $3$-D projection of F-ratio values in AF-MF plane. Different auditory and modulation bins show varying discriminative characteristics for different emotions. For every emotion class, F-ratio peaks are observed at low MF bins showing their potential for emotion discrimination. Similarly, low AF also shows high F-ratio for every class except \emph{Sad}. High arousal emotions (\emph{Anger}, \emph{Happy}, \emph{Fear, etc.}) in general show greater F-ratios at high MF bins as compared to low arousal emotions (\emph{Sad}, \emph{Boredom}). The highest F-ratio value with respect to \emph{Neutral} is observed for \emph{Anger} and lowest for \emph{Boredom} emotion class. For \emph{Anger} and \emph{Happy} classes, low AF bins exhibit higher discriminative characteristic. \emph{Disgust} shows F-ratio peaks over wide range of AF and corresponding MF bins. In \emph{Fear}, F-ratio peaks are observed at low and high AF values with gradual slope towards increasing MF bins. The presence of moderately higher F-ratio values at MF bins is a result of increase in speaking rate in high arousal emotions. 
\emph{Boredom} class has lowest F-ratio values, mostly focused at low AF and low MF bins, whereas, \emph{Sad} shows higher discrimination w.r.t. \emph{Neutral} over low and high AF and MF bins. Lower F-ratio values for \emph{Boredom} also indicate its similarity in characteristics with \emph{Neutral} class. The F-ratio analysis again shows the higher discrimination potential of joint AF and MF bins, with respect to \emph{Neutral} emotion.

\subsection{Comparison between CQT-MSF and Scattering Transform}
\label{comp_scat_msf}

Our proposed CQT-MSF feature, combined with standalone CQT, has striking similarity with \emph{scattering transform} feature representation of $1$-D signals. Authors in \cite{anden} compute scattering coefficients of $1$-D signals and show their characteristic invariance against temporal shifts and deformations. The features (or coefficients) are computed by convolving the signal with a set of predefined filter kernels. The feature extraction process includes the following steps: 1) Scalogram computation by passing the signal through a bank of wavelet filters. 2) Passing the obtained time-series of frequency bins in scalogram through another set of wavelet filterbank to obtain modulation spectrogram. 3) Introduce stability to deformations by low-pass filtering the signal, scalogram and modulation spectrogram coefficients. The scattering transform coefficients are mathematically described as,

\begin{equation}
    S_{J_2} x(t) = U_2 x*\phi_{2^J}(t) = \int U_2 x(u)\phi_{2^J}(t-u)du
\end{equation}

\noindent where,
\begin{equation}
    U_2 x = U[\lambda_2]U[\lambda_1]x = ||x*\psi_{\lambda 1}|*\psi_{\lambda 2}|.
\end{equation}

Here, $x$ is the $1$-D signal, $\phi_{2^J}(t)$ defines the averaging low-pass filter with scale $2^J$, $\psi_{\lambda_N}$ describes the $N$th layer complex \emph{Morlet} wavelet filterbank (layer $1$ are scalogram coefficients and layer $2$ constitutes modulation coefficients) and operator `$*$' is the convolution operator. Scattering coefficients are found useful in various speech and audio processing domains, e.g., speech recognition \cite{anden}, speaker identification \cite{ghezaiel:hal-03086433}, urban and environmental sound classification \cite{bauge2013}, etc. In \cite{singh2021deep}, scattering coefficients also showed improvement in SER performance over mel-frequency cepstral coefficients (MFCCs).\\

\renewcommand{\thesubfigure}{\alph{subfigure}}
\begin{figure}[t]
    \centering
    \begin{subfigure}[t]{0.5\textwidth}
        \hbox{\hspace{-0.7cm}\includegraphics[scale=0.6]{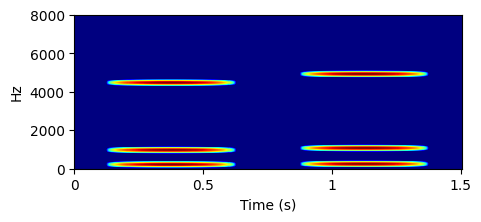}}\caption{STFT}
    \end{subfigure}%
    \begin{subfigure}[t]{0.5\textwidth}
        \includegraphics[scale=0.6]{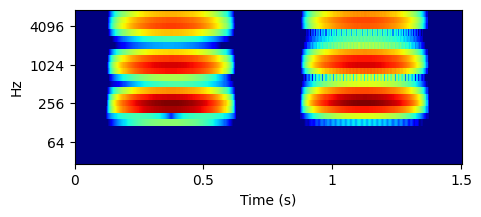}\caption{CQT}
    \end{subfigure}%
    \vspace{0.75cm}
    \begin{subfigure}[t]{0.6\textwidth}
        \includegraphics[scale=0.6]{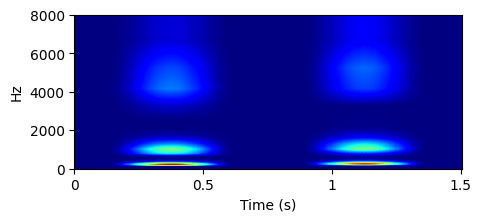}
        \caption{CQT with linear frequency scale}
    \end{subfigure}
    \caption{Visual description of deformation stability of STFT and CQT. a), Signal $x(t)$ (left) and its deformed version $x`(t)$ (right). b), CQT representation of the same signal and its deformed version. c) CQT of the original and deformed signal projected on linear frequency scale. Figure taken with permission from~\cite{SINGH2022103712}.}
    \label{deform_stable}
\end{figure}

In our proposed CQT-MSF feature, the CQT time-frequency representation is similar to the first layer scalogram coefficients computed by scattering transform. Similarly, the MSF computed over CQT is similar to the modulation spectrogram computed over scalogram in second layer of scattering transform. Also, CQT follows the same constant-Q non-linearity as followed by the filterbanks in both first and second layers of the scattering transform. However, the averaging performed in scattering coefficients to obtain invariance to time-shift and deformations is absent in CQT-MSF. The design parameter (e.g., bandwidth) of the low-pass filter which performs this averaging in scattering transform is manually selected depending upon the input signal characteristics. To address the absence of time-shift invariance, we employ a $2$-D convolutional neural network over the computed CQT-MSF. The employed CNN architecture includes multiple layers with different filter scale values in different layers. According to \cite{wiatowski}, convolutional neural networks inherently exhibit invariance in vertical direction (direction of network depth) which mainly appears due to feature pooling. Hence, a generic CNN architecture with pooling layers can learn to apply the required averaging and obtain the required time-shift invariance characteristic.\\

Regarding sensitivity to deformation, authors in \cite{wiatowski} and \cite{philipp} prove that convolutional feature extractors provide inherent but limited deformation stability. The extent of this stability depends upon the deformation sensitivity of the input signal. Signals which are slowly varying or band-limited are more deformation insensitive than signal with sudden changes or discontinuities \cite{philipp}. As the CQT also provides a non-uniform filterbank representation as provided by mel-filters, similar stability to temporal deformation can be assumed in CQT filterbank as well. Fig.~\ref{deform_stable} shows the deformation stability of STFT, CQT and CQT with linear frequency scale for a signal $x(t)$ deformed by a factor $\epsilon t$ (i.e, $x'(t) = x(t-\epsilon t) = x((1-\epsilon)t)$) \cite{anden}. The upward shift of spectral response in STFT appears due to the `$\epsilon t$' term, leading to instability to deformation imposed by $\epsilon t$. However, in CQT, spectral responses of deformed signal do not show any major frequency shift. Instead, the deformed signal well-overlaps with the original signal because of higher filter bandwidth at higher frequencies. This shows that CQT is indeed deformation stable as compared to STFT. The linear frequency CQT plot is given for comparison of STFT and CQT over linear frequency scale, hence confirming the deformation stability of CQT in both linear and non-linear frequency scales. \\

Therefore, convolutional neural network layers can be used to inherently provide the required time-shift and deformation invariance for better emotion-rich representation at its output. We hence write the feature extracted by convolution layers of our employed DNN model as,

\begin{equation}
    S = F(||x*\psi_{\lambda1}|*\psi_{\lambda2}|)
\end{equation}

\noindent where, $F(.)$ is the function estimated by $2$-D convolution layers, and $\psi_{\lambda1}$ and $\psi_{\lambda2}$ corresponds to the filterbanks $h_{\omega_n^a}(t)$ and $g_{\omega_k^m}(t)$ used in the CQT-MSF generation (Fig.~\ref{CQT-MSF filterbanks}). Another point of dissimilarity is the difference between the basis functions used in the filterbank of scattering transform and CQT-MSF. The former uses \emph{Morlet} wavelets, whereas, the latter employs sinusoids multiplied with \emph{Hann} window function. The ripples observed in the frequency response of the filters in Fig.~\ref{CQT-MSF filterbanks} is because of the small spectral leakage in the \emph{Hann} window.

\section{Experimental Setup}
\label{exp_setup}

\subsection{Database Description}
\label{data_used}
For analysis of CQT-MSF and its comparison with mel-scale features, we perform experiments with two different speech corpora. We use Berlin EmoDB and RAVDESS datasets which are most widely used and publicly available.

\subsubsection{Berlin Emotion Database (EmoDB)}
Berlin Emotion Database \cite{burkhardt2005database} contains acted emotional speech recordings of $10$ professional artists ($5$ female and $5$ male). The actors speak ten emotionally neutral and phonetically rich sentences in German language. Seven different emotion categories are used in the database: \textit{Anger, Happy, Fear, Sad, Boredom, Disgust,} and \textit{Neutral}. To evaluate the authenticity of recordings, listening test was performed by $20$ subjects. A total of $800$ utterances were recorded but only $535$, having more than $80\%$ recognition rate and $60\%$ naturalness, were finally selected. Our choice of this database is explained by its diligent recording setup, popularity in SER domain \cite{eyben2015geneva, zhang2017speech, wu2011automatic, mao2014learning,  bitouk2010class,  wang2015speech, deb2018multiscale, Ntalampiras2012} and its free availability.

\subsubsection{Ryerson Audio-Visual Database of Emotional Speech and Song (RAVDESS)}
The RAVDESS database \cite{livingstone2018ryerson} contains acted utterances of $12$ male and $12$ female artists speaking English language. A total of $7536$ clips were recorded in three different modalities, namely audio-only, video-only, and audio-video, out of which the audio-only modality contains $1440$ spoken utterances from all speakers. The database includes eight different emotion categories (\textit{Happy, Anger, Sad, Neutral, Disgust, Calm, Surprised,} and \textit{Fear}) with two intensity levels, strong and normal. Recorded clips were evaluated by $319$ subjects out of which $247$ tested the validity and $72$ evaluated test-retest reliability of recordings. An average of $60\%$ accuracy was obtained in validity test over recordings of all emotions. Up-to-date design and inclusion of an extensive emotion set with varying intensities make this an important database for SER.\\

To further increase the diversity in training data, five-fold data augmentation following the x-vector \emph{Kaldi}\footnote{\url{https://github.com/kaldi-asr/kaldi/tree/master/egs/voxceleb/v2}}~recipe is used \cite{snyder2018x}. The augmented data involves adding additive and reverberation noises over clean speech samples. The RAVDESS database is downsampled to $16$~kHz before data augmentation and feature extraction.

\subsection{Parameter settings for feature extraction}
We compare the performance of proposed CQT-MSF features with baseline CQT and MFSC. The different parameter values used for CQT and MFSC are based on our preliminary comparison of the two methods~\cite{singh2021}. For CQT computation, we select the minimum frequency value $F_{\mathrm{min}}$ to be $32.7$~Hz and $F_{\mathrm{max}}$ equal to the Nyquist frequency. This provides a total of eight octaves over complete frequency range. Every frequency octave contains three bins which provides a total of $24$ frequency bins over complete frequency range ($F_{\mathrm{min}}$ to $F_{\mathrm{max}}$). The obtained CQT representation corresponds to $24$-channel early auditory stage feature extraction. Another important parameter in CQT computation is hop length which is the number of samples the observation window advances between successive frame-shifts. We keep the hop length value fixed at $64$. For cortical analysis, the above-mentioned CQT representation is passed through another set of constant-Q filterbank, referred to as modulation filterbank, as described in Section~\ref{mod_spec_section}. We use an 8-channel modulation filterbank with center frequencies ranging from $0.5$ to $64$~Hz covering a total of eight frequency octaves.\\

For a fair comparison with CQT and CQT-MSF, similar modification in parameter values of MFSC is applied. We use $24$-filter bank for MFSC computation over STFT with $512$ frequency bins. The frame size is fixed to $320$ samples with hop length of $64$ samples over $16$~kHz sampling frequency. We also compute the modulation spectrum coefficients by using MFSC time-frequency representation for comparison with CQT-MSF features. We refer to these as MFSC-MSF features. We use the same \emph{LibROSA} toolkit for MFSC feature generation.

\subsection{Evaluation Methodology}
\label{eval_metho}

Unlike other speech processing tasks, such as automatic speech recognition (ASR), automatic speaker verification (ASV), the SER lacks standardization of evaluation methodology for performance benchmarking on publicly available datasets. This leads to a wide variation in results across different works reported in the literature. Some examples of differences in evaluation protocol are the use of some selected emotions from the databases, the choice of performance evaluation metric, the selection of cross-validation strategy, the differences in the selected emotion classes, etc. Because of these reasons, meaningful comparison of obtained results with those reported in the literature becomes inaccurate, if not impossible, in SER research. \\

We adopt a leave-one-speaker-out (LOSO) cross-validation strategy for evaluation and benchmarking. The databases are divided into train/validation/test groups with every group containing disjoint speakers. The test and validation group contain utterances from one speaker and the remaining speakers are kept for training. This keeps the total number of train/validation/test sets equal to the number of speakers in every database. The final performance over a database is reported by averaging the performance metrics obtained for every train/validation/test group. For SER, speaker-dependent testing is known to fair better than speaker-independent testing \cite{schuller2005speaker}. However, speaker-independent sets of the database eliminate the chances of the trained classifier being biased towards a set of speakers and also simulates the real-world scenario in a better way. Although LOSO cross-validation is computationally expensive, due to small database sizes in SER, the increase in complexity can be safely ignored. Also, keeping a single speaker for testing allows more training data to be available which is essential with small databases. \\

\begin{table}[t!]
\centering
\caption{The parameters of CNN architecture for SER. The number of $2$D-Conv layers and the kernel sizes are inspired from the x-vector TDNN architecture~\cite{snyder2018x}. Maxpooling applied after every $2$D-Conv layer provides time and frequency invariant feature representations.}
\begin{adjustbox}{width=0.575\columnwidth,center}
\renewcommand{\arraystretch}{1.4}
\begin{tabular}{cccc}
\hline
\textbf{Layer} & \begin{tabular}[c]{@{}c@{}} \textbf{No. of} \\ \textbf{Filters} \end{tabular} & \begin{tabular}[c]{@{}c@{}} \textbf{Height} \\ \textbf{(Frequency)}\end{tabular} & \begin{tabular}[c]{@{}c@{}} \textbf{Length} \\ \textbf{(Time)}\end{tabular} \\ \hline
\hline
2-D Conv & 128 & 5 & 5 \\ 
Maxpool & - & 2 & 1 \\
2-D Conv & 128 & 3 & 3 \\
Maxpool & - & 2 & 1 \\
2-D Conv & 128 & 3 & 3 \\
Maxpool & - & 2 & 1 \\
2-D Conv & 128 & 1 & 1 \\
Maxpool & - & 2 & 1 \\
\begin{tabular}[c]{@{}c@{}} Global Average \\ Pool (GAP) \end{tabular}  & - & - & - \\
Fully Connected  & 64 & - & - \\
Softmax & \#Classes & - & - \\
\hline
\end{tabular}
\end{adjustbox}
\label{cnn_param_tab}
\end{table}


\subsection{Classifier Description}
\label{classifier_desc_sec}
In this work, we use two different machine learning frameworks to evaluate performances of studied features: (1)~convolutional neural network with fully connected layer for emotion classification (termed henceforth as DNN). (2)~Convolutional layers to extract emotion embeddings and SVM to classify embeddings into emotion classes (termed as DNN-SVM). Our selection of these is inspired from the success of embeddings based networks~\cite{zhang2017speech, snyder2018x, dawalatabad21_interspeech} and fully DNN-based frameworks~\cite{peng2020} in speech processing. Performance evaluation over these also enables us to compare the SER efficiency of the two DNN frameworks. \\

The DNN-SVM framework comprises of two parts: embedding extraction from convolutional layers of the trained model used in the DNN framework, and final classification using SVM. The SVM is trained and tested over the embeddings extracted from the trained DNN model.
We extract embeddings at the output of global average pooling (GAP) layer, placed after the final convolutional layer. These embeddings are processed with an SVM back-end for performance evaluation. In SVM model, we empirically select the value of regularization parameter $C$ and the expanse/width of the \emph{radial basis function} kernel (parameter $\gamma$) to $1$ and $0.001$ respectively~\cite{libsvm}.  \\ 

In the DNN framework, to train and validate the model, the speech utterances from every database are chunked into segments of $100$ frame length with $50\%$ overlap across consecutive frames. With $64$ samples hop and $16$~kHz sampling rate, this corresponds to $400$~ms speech duration. Our choice of $400$~ms is based on the reports in SER literature which explain that segment length greater than $250$~ms contains required information for emotion prediction~\cite{zhang2017speech}. However, for testing, complete utterances are used to test model performance. This is done as the labels provided to emotion speech recordings are over complete utterances and not over segments. This approach also leads to increase in available data samples for training. Similarly, in DNN-SVM framework, the train embeddings are generated over segments of speech utterances with $100$ frame length ($400$~ms), whereas, test embeddings are generated from complete utterances. \\

Table~\ref{cnn_param_tab} describes the DNN architecture employed in this work. We use cross entropy optimizer with learning rate value of $0.001$ with $64$ batch size and dropout value of $0.3$ applied over only the fully connected (FC) layer. The model is trained for $50$ epochs and the version with the best performance on validation set is used for testing.  \\

\begin{figure}[t]
    \begin{subfigure}[t]{0.55\textwidth}
        \hspace{-0.25cm}
        \includegraphics[scale=0.45]{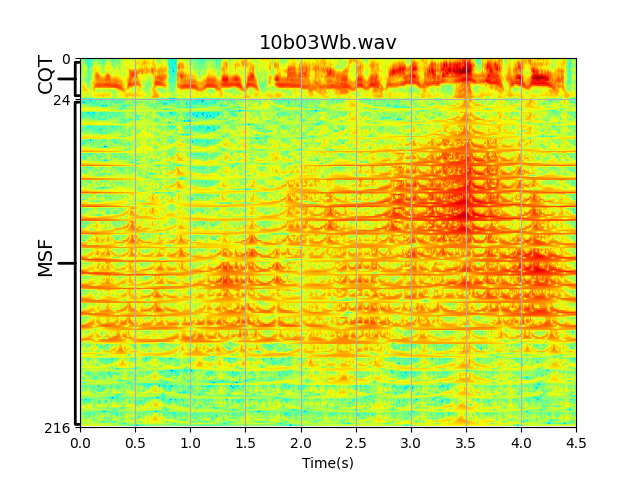}
        \caption{CQT-MSF of \emph{Anger} emotion}
    \end{subfigure}%
    \begin{subfigure}[t]{0.55\textwidth}
        \hspace{-0.25cm}
        \includegraphics[scale=0.45]{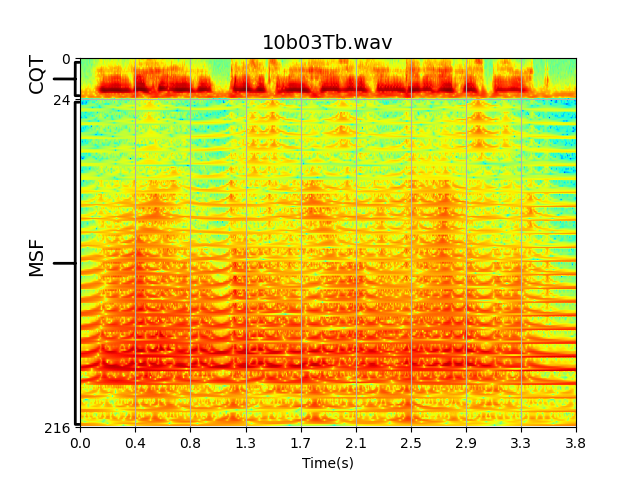}
        \caption{CQT-MSF of \emph{Sad} emotion}
        \label{cqt_msf_example_figure}
    \end{subfigure}%
    \caption{Logarithm of feature fusion of CQT and modulation spectral features, i.e., CQT-MSF, extracted from CQT over utterances taken from EmoDB database. The first $24$ bins on y-axis correspond to the CQT spectrogram. Bins that follow include stacking of $8$ modulations bins corresponding to every auditory bin. The total number of bins then becomes, $24$ auditory bins $+$ $24$ auditory bins $\times~8$ modulation bins for every auditory bin$~=~24~$auditory bins$~+~192~$MSF bins$~=~216$ total bins on y-axis).}
    \label{CQT-MSF-plot}
\end{figure}

As the convolution layers accept input in $2$-D (time-frequency) form, we use two different strategies to combine CQT/MFSC time-frequency representation with their MSF features. In the first method, we directly concatenate the CQT/MFSC with its corresponding MSF features over the frequency axis. This leads to a representation with time frames in x-axis and early auditory frequency bins, followed by modulation frequency bins corresponding to every auditory bin, placed in succession over y-axis. Fig.~\ref{CQT-MSF-plot} shows the $2$-D representation obtained with concatenation of CQT/MFSC with the corresponding MSF. 
In second approach, to better combine the information from time-frequency and modulation features, we use an embedding fusion based DNN architecture. The architecture consists of two parallel but similar branches of convolutional and GAP layer, followed by a common FC and softmax layer. For both feature fusion and embedding fusion, the embeddings are extracted from the GAP layer of DNN model.

\subsection{Evaluation Metrics}

For performance evaluation, we use accuracy and UAR metrics. We chose these metrics owing to their popularity in SER and also for better comparison of results with the literature. Accuracy is defined as the ratio between the number of correctly classified utterances to the total number of utterances in test set. According to \cite{rosenberg2012classifying}, the UAR metric is given as:

\begin{equation}
    \mathrm{UAR} = \frac{1}{K} \sum_{i=1}^{K} \frac{A_{ii}}{\sum_{j=1}^{K} A_{ij}}
\end{equation}
 
\noindent here, $A$ is called the contingency matrix, $A_{ij}$ refers to the number of samples in class $i$ classified as class $j$, and $K$ is the total number of classes. As accuracy is considered \emph{unintuitive} for databases with uneven samples across different classes, we use UAR to measure the validation set performance of the DNN model and to select the best performing model over the set of epochs.


\section{Results \& Discussion}
\label{res_and_dis}

\subsection{Performance Comparison of Different Features}

\renewcommand{\thesubfigure}{\alph{subfigure}}
\begin{figure}[t]
    \centering
    \begin{subfigure}[t]{0.55\textwidth}
        \hbox{\hspace{-0.4cm}\includegraphics[scale=0.23]{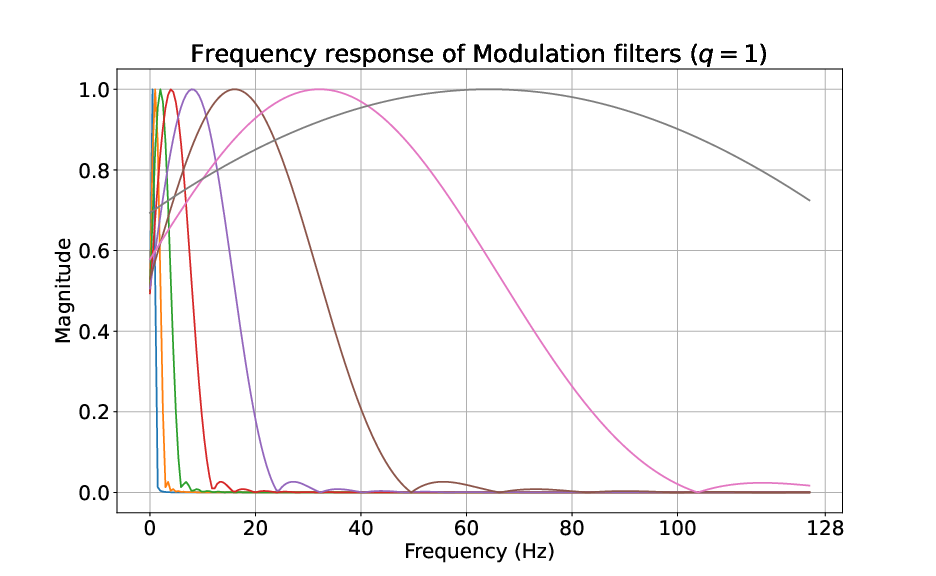}}\caption{Filter scale $q~=~1$}
    \end{subfigure}%
    \begin{subfigure}[t]{0.5\textwidth}
        \hbox{\hspace{-0.4cm}\includegraphics[scale=0.23]{Fig2_2_Mod_filt_scale2.png}}\caption{Filter scale $q~=~2$}
        \label{Mod_filts_2}
    \end{subfigure}%
    \caption{Modulation filter banks for different values of filter scale factor $q$. Filters with $q~=~1$ have same center frequencies but higher bandwidth than filters with $q~=~2$.}
    \label{Mod_filts}
\end{figure}

\begin{table}[t]
\centering
\caption{Comparison between different filter scale ($q$) values of modulation filters for experiments performed with feature-fused CQT-MSF over EmoDB database. Given values are in percentages.}
\renewcommand{\arraystretch}{1.3}
\begin{tabular}{ccccccc}
\hline
\multirow{2}{*}{\textbf{\begin{tabular}[c]{@{}c@{}} Classification \\ Framework\end{tabular}}} & \multicolumn{2}{c}{\textbf{$\boldsymbol{q~=~1}$}} & \multicolumn{2}{c}{\textbf{$\boldsymbol{q~=~2}$}}  & \multicolumn{2}{c}{\textbf{$\boldsymbol{q~=~3}$}}  \\ 
 & \multicolumn{1}{c}{\textbf{Accuracy}} & \textbf{UAR} & \multicolumn{1}{c}{\textbf{Accuracy}} & \textbf{UAR}  & \multicolumn{1}{c}{\textbf{Accuracy}} & \textbf{UAR} \\ \hline
 \hline
\begin{tabular}[c]{@{}c@{}} DNN\end{tabular} & \multicolumn{1}{c}{76.97} & 68.25 & \multicolumn{1}{c}{70.79} & 64.93  & \multicolumn{1}{c}{69.77}  & 64.27 \\ 
\begin{tabular}[c]{@{}c@{}} DNN-SVM\end{tabular} & \multicolumn{1}{c}{79.86} & 76.17 & \multicolumn{1}{c}{79.50} & 77.00  & \multicolumn{1}{c}{74.28}  & 70.91 \\ \hline
\end{tabular}
\label{Qvalue_analysis}
\end{table}

\begin{table}[t!]
\centering
\renewcommand{\arraystretch}{1.4}
\caption{Performance comparison of combined early auditory and cortical features over EmoDB database. Filter scale ($q$) value of $1$ is used in modulation filterbank for MSF computation. Given values are in percentages.}
\begin{tabular}{ccccc}
\hline
\multirow{2}{*}{\textbf{Features}} & \multicolumn{2}{c}{\textbf{DNN}} & \multicolumn{2}{c}{\textbf{DNN-SVM}} \\ 
 & \multicolumn{1}{c}{\textbf{Accuracy}} & \multicolumn{1}{c}{\textbf{UAR}} & \multicolumn{1}{c}{\textbf{Accuracy}} & \textbf{UAR} \\ \hline
 \hline
\begin{tabular}[c]{@{}c@{}}CQT-MSF \\ (Feature Fusion)\end{tabular} & \multicolumn{1}{c}{76.97} & 68.25 & \multicolumn{1}{c}{79.86} & 76.17 \\ 
\begin{tabular}[c]{@{}c@{}}CQT-MSF\\ (Embedding Fusion)\end{tabular} & \multicolumn{1}{c}{77.99} & 71.74 & \multicolumn{1}{c}{79.32} & 75.48 \\ 
\begin{tabular}[c]{@{}c@{}}MFSC-MSF\\ (Feature Fusion)\end{tabular} & \multicolumn{1}{c}{61.45} & 55.23  & \multicolumn{1}{c}{69.02} & 64.45 \\ 
\begin{tabular}[c]{@{}c@{}}MFSC-MSF\\ (Embedding Fusion)\end{tabular} & \multicolumn{1}{c}{60.80} & 57.92 & \multicolumn{1}{c}{67.74} & 64.89 \\ \hline
\end{tabular}
\label{msf_analysis}
\end{table}

\begin{table}[t!]
\centering
\renewcommand{\arraystretch}{1.4}
\caption{Performance comparison of early auditory and cortical features taken separately over EmoDB database. Filter scale ($q$) value of $1$ is used in modulation filterbank for MSF computation. Given values are in percentages.}
\begin{tabular}{ccccc}
\hline
\multirow{2}{*}{\textbf{Features}} & \multicolumn{2}{c}{\textbf{DNN}} & \multicolumn{2}{c}{\textbf{DNN-SVM}} \\ 
 & \multicolumn{1}{c}{\textbf{Accuracy}} & \multicolumn{1}{c}{\textbf{UAR}} & \multicolumn{1}{c}{\textbf{Accuracy}} & \textbf{UAR} \\ \hline
\hline
\begin{tabular}[c]{@{}c@{}}MSF \\ (Computed over CQT)\end{tabular} & \multicolumn{1}{c}{72.67} & 66.32 & \multicolumn{1}{c}{78.73} & 76.33 \\ 
\begin{tabular}[c]{@{}c@{}}MSF \\ (Computed over MFSC)\end{tabular} & \multicolumn{1}{c}{59.35} & 53.44 & \multicolumn{1}{c}{65.24} & 60.54 \\ 
CQT & \multicolumn{1}{c}{71.77} & 64.91 & \multicolumn{1}{c}{76.74} & 73.21 \\ 
MFSC & \multicolumn{1}{c}{61.62} & 57.32 & \multicolumn{1}{c}{66.18} & 62.58 \\ \hline
\end{tabular}
\label{feat_analysis}
\end{table}

\begin{table}[h!]
\caption{Feature performance comparison over different databases with DNN-SVM framework. Given values are in percentages}
\centering
\label{data_comp}
\renewcommand{\arraystretch}{1.4}
\begin{adjustbox}{width=\columnwidth, center}
\begin{tabular}{@{}ccccccccc@{}}
\hline
\multirow{3}{*}{\textbf{Database}} & \multicolumn{2}{c}{\textbf{MFSC}} & \multicolumn{2}{c}{\textbf{CQT}} & \multicolumn{2}{c}{\begin{tabular}[c]{@{}c@{}} \textbf{CQT-MSF} \\ \textbf{(Feature Fusion)}\end{tabular}} & \multicolumn{2}{c}{\begin{tabular}[c]{@{}c@{}} \textbf{Chance level} \\ \textbf{performance}\end{tabular}}\\ 
 & \textbf{Accuracy} & \textbf{UAR} & \textbf{Accuracy} & \textbf{UAR} & \textbf{Accuracy} & \textbf{UAR} & \multicolumn{2}{c}{\textbf{Accuracy}} \\ \hline
 \hline
EmoDB & 66.01 & 61.75 & 76.74 & 73.21 & 79.86 & 76.17 & \multicolumn{2}{c}{14.28}\\ 
RAVDESS & 36.94 & 36.16 & 48.68 & 44.64 & 52.24 & 48.83 & \multicolumn{2}{c}{12.5}\\ \hline

\end{tabular}
\end{adjustbox}
\end{table}

First, we experimentally optimize the constant-Q filters used for CQT-MSF computation by varying the value of filter scaling factor ($q$).

Fig.~\ref{Mod_filts} shows the frequency response of modulation filter banks with $q$~=~$1$ and $2$. Since $q$ affects the time resolution of filters as given in Eq.~\ref{eq:scale}, filters with $q$~=~$1$ are wider (have higher bandwidth) which leads to clipping of the filter frequency response at low frequencies. This leads to inclusion of zero-frequency (or DC) components as well. Also, because of greater overlap between filters, the generated filter outputs have higher redundancy. Increased redundancy helps convolutional layers to better extract required emotion correlation among modulation frequency bins. With $q$~=~$2$, the filter responses remain limited inside the frequency range providing a less redundant filterbank structure as shown in Fig.~\ref{Mod_filts_2}. Table~\ref{Qvalue_analysis} reports the difference in results obtained for modulation features computed with different values of scaling factor $q$. Filterbank structure with $q$~=~$1$ outperforms the arrangement with $q$~=~$2$ and $3$. Hence, we select modulation filters with $q$~=~$1$ for further experiments. We perform optimization and detailed experimentation of $q$ over only EmoDB database due to its small size and similar trends with other databases.\\

\begin{figure}
    \centering
    \begin{subfigure}[h]{0.5\textwidth}
        \hbox{\hspace{0cm}\includegraphics[scale=0.2]{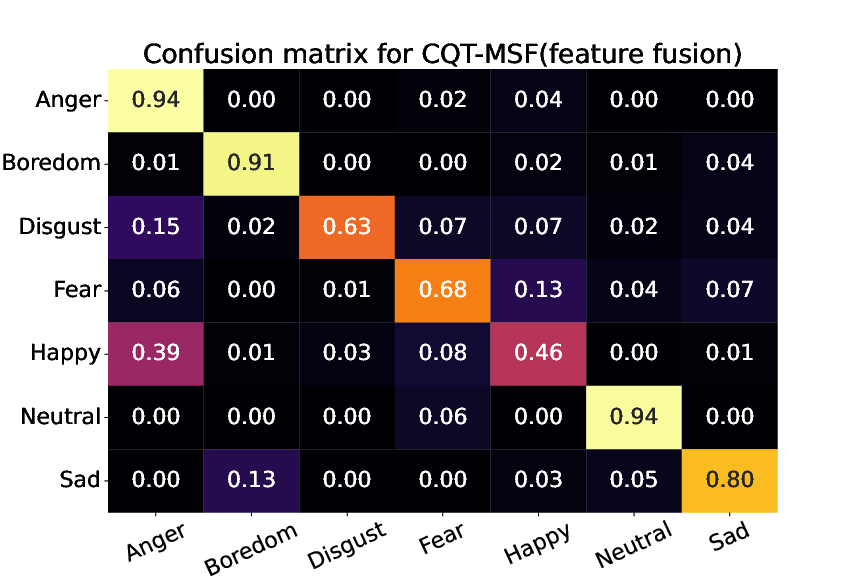}}
    \end{subfigure}%
    \begin{subfigure}[h]{0.5\textwidth}
        \hbox{\hspace{0cm}\includegraphics[scale=0.2]{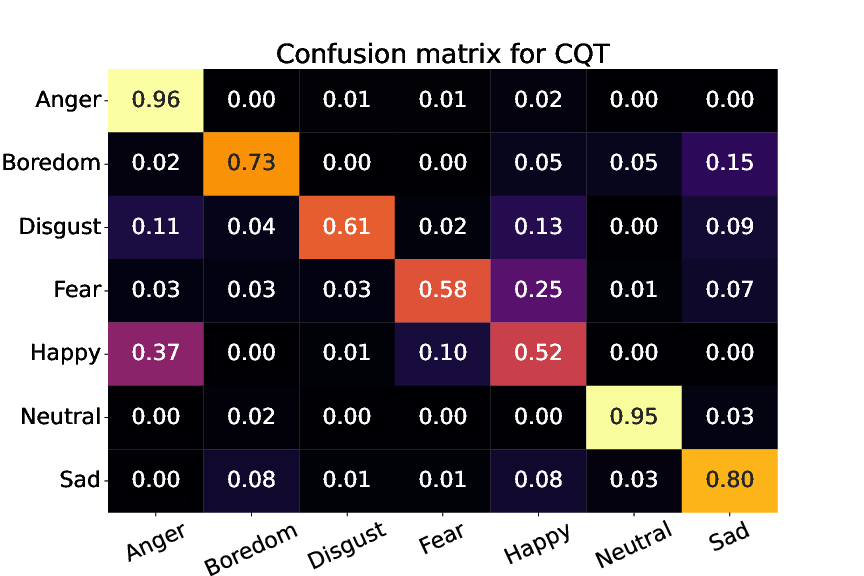}}
    \end{subfigure}
    \begin{subfigure}[h]{0.5\textwidth}
        \hbox{\hspace{0cm}\includegraphics[scale=0.2]{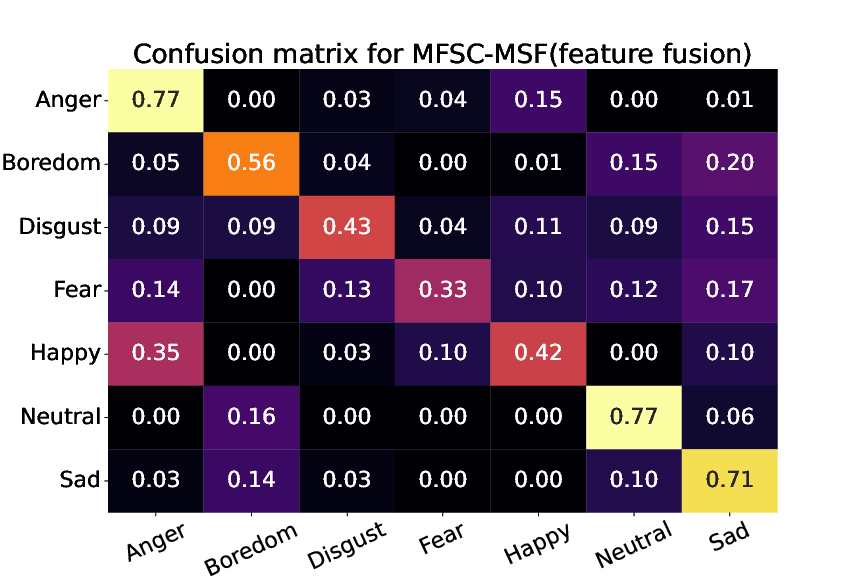}}
    \end{subfigure}%
    \begin{subfigure}[h]{0.5\textwidth}
        \hbox{\hspace{0cm}\includegraphics[scale=0.2]{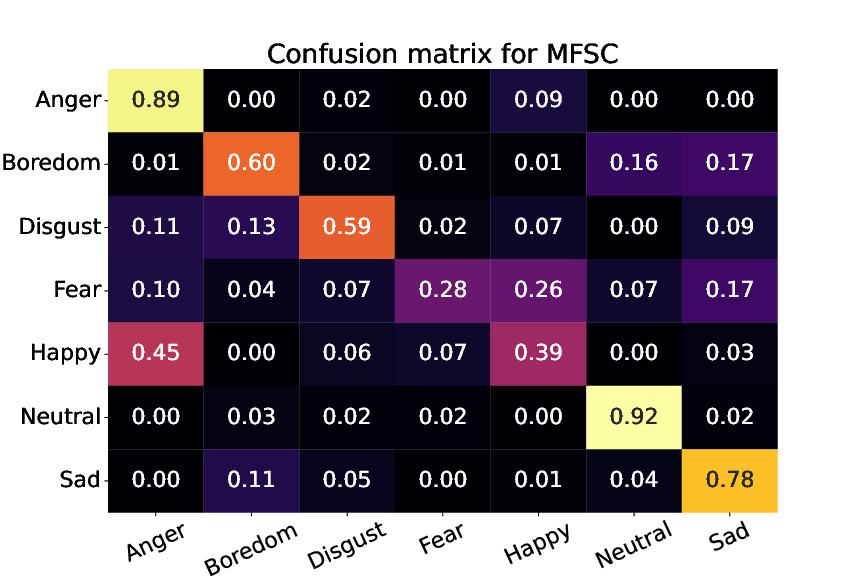}}
    \end{subfigure}
    \caption{Confusion matrices for CQT-MSF (feature fusion), MFSC-MSF (feature fusion), CQT, and MFSC features. CQT-MSF shows best comparative performance in classifying different emotion classes of EmoDB database. Matrices are computed over DNN-SVM framework owing to its greater performance.}
    \label{conf_mats}
\end{figure}

Table~\ref{msf_analysis} shows the performance of time-frequency representations combined with their corresponding modulation features for different classification frameworks over EmoDB database. The DNN-SVM classification framework outperforms DNN framework for every feature and over both performance metrics. This observation is counter-intuitive as the activations extracted from the same convolutional layer in DNN-SVM and DNN framework, end up performing better with SVM at back-end but not with fully-connected layer at the back-end. Regarding the two different feature fusion types, there is no specific pattern, in terms of performance improvement, among the classification frameworks. With CQT-MSF, feature fusion outperforms embedding fusion over DNN framework, whereas, embedding fusion outperforms feature fusion for DNN-SVM framework. For MFSC-MSF, the trend is opposite to that of CQT-MSF fusion results. However, CQT-MSF in both fusion styles performs better than MFSC-MSF. \\

To further analyse the contribution of early auditory and cortical features taken separately over SER, we perform experiments to analyse the performance of standalone MSF extracted over CQT/MFSC (without any type of fusion). From the results in Table~\ref{feat_analysis}, cortical features (standalone MSF over CQT) outperforms standalone CQT. However, the same is not true for mel-scale based features. The MSF computed over MFSC shows poor performance as compared to standalone MFSC. This questions the usability of temporal trajectories of the mel-scale time-frequency representation for emotion classification. The inferiority of MFSC against CQT is also indicated by direct comparison between CQT and MFSC. CQT outperforms MFSC in both DNN and DNN-SVM classification frameworks. Among different CQT feature types, MSF computed over CQT assumes very similar performance in contrast to both fusion types of CQT-MSF. This phenomenon describes the higher emotion relevance of the temporal modulations of the low-frequency regions emphasised by CQT.    \\

Fig.~\ref{conf_mats} shows the confusion matrices for different features over EmoDB database. We choose only the feature-fused CQT-MSF and MFSC-MSF for comparison, owing to their improved performance as compared to embedding fusion in DNN-SVM framework. Even though CQT-MSF is comparatively better at classifying different emotion classes, some instances of \emph{Happy} and \emph{Disgust} are confused with \emph{Anger}, and that of \emph{Fear} are confused with \emph{Happy}. The highest misclassification is in \emph{Happy-Anger} emotion pair which have similar arousal but opposite valence characteristic. 
This observation is found to be consistent among various SER works~\cite{zhang2017speech, peng2020, deb2019} and can also be related to the similar F-ratio characteristics of \emph{Anger} and \emph{Happy} in Fig.~\ref{f_ratio_modspec}. Among low-arousal classes, some confusion in \emph{Sad-Boredom} is also visible in CQT-MSF. As prosody features are less effective in valence discrimination~\cite{iker2010}, the confusion among pairs with similar arousal but opposite valence characteristics can be attributed to higher emphasis over pitch in constant-Q scale based spectral representation. However, modulations computed across constant-Q scale reduce this confusion, which is evident from the comparison between standalone CQT and CQT-MSF. In standalone CQT, confusion among both low and high arousal emotions is higher and appears among multiple emotion classes (e.g., \emph{Disgust} and \emph{Fear} with \emph{Happy}). In MFSC and MFSC-MSF, as compared to CQT-based features, the misclassification is more prominent across multiple classes. Unable to emphasise speech prosody, the emotion classification ability of MFSC over arousal scale is inferior to CQT-based features (e.g., increased confusion of \emph{Boredom} with \emph{Neutral}, and \emph{Fear} with \emph{Sad}). Also, modulations of MFSC are computed with more focus on high and mid speech frequencies and less focus on prosody at low frequencies, further deteriorating the performance.  \\

At the end, we compare the SER performance of CQT-MSF feature with the scattering transform coefficients. As scattering network is also a deep convolutional network, its comparison with our proposed CQT-MSF with DNN-SVM classification framework shows the superiority of automatic feature extraction and required time and frequency-shift invariance learning for SER. The scattering coefficients are computed using the similar train/test strategy. The training speech utterances are chunked to $400$~ms segments with $50$\% overlapping, whereas testing utterances are used as is. Following our experiments in~\cite{singh2021deep}, the Q-factor value (Q) for first layer coefficients is chosen as Q~=~$5$. As the training segment size is fixed to $400$~ms ($6400$ samples at $16$~kHz), the maximal wavelet length or averaging scale ($T$) is kept $4096$ samples for both training and validation. However, since we use complete utterances in test, the duration $N$ for testing is empirically fixed to $51000$ samples ($3.18$ seconds at $16$~kHz). Longer utterances are chopped to contain only $51000$ samples, whereas shorter frames are zero-padded. We obtain \boldsymbol{$72.67\%$} accuracy and \boldsymbol{$69.8\%$} UAR with scattering coefficients outperforming MFSC, and indicating the requirement of time and deformation stability in SER. However, the performance is inferior to that with the proposed CQT-MSF (especially with feature-fusion). The superiority of CQT-MSF shows that deep networks learn to provide better time and deformation stability, as described in Section~\ref{comp_scat_msf}, while extracting the emotion relevant information from multiple convolution layers. Although scattering transform also involves convolutions and averaging for stability, it is performed using fixed kernels which are not automatically learned/optimized to improve performance.\\

Table~\ref{data_comp} shows the results obtained with different features over EmoDB and RAVDESS databases. The proposed CQT-MSF feature outperforms other features over RAVDESS database as well. This explains the suitability of CQT-MSF features or two stage auditory analysis for SER over different databases. Compared to EmoDB, the relative performance improvement with CQT-MSF, CQT over MFSC is higher in RAVDESS database.  \\

\subsection{Comparison with Related Works}

In this subsection, we compare our obtained results with related works in SER. Among SER literature, different strategies are used to evaluate system performances, for example, use of different databases, number of emotion classes used in the databases, different evaluation methodologies, etc. These differences make direct comparison of SER works difficult. The evaluation methodology, in terms of, cross-validation scheme, train/test split, speaker dependent/independent testing, etc. are found to differ substantially in SER literature. Lack of reproducible research in SER domain also leads to uncertainty, leading to difficulty in comparison. Hence, a comparison made with other relevant SER literature can not be considered accurate. \\ 

Due to the above mentioned issues, to justify our obtained results, we implement different studies from the literature by using our proposed CQT-MSF feature and experimental framework. Table~\ref{comp_res_table} shows the list of selected works and the corresponding performances obtained. Section~\ref{eval_metho} and~\ref{classifier_desc_sec} of the manuscript describe the experimental framework employed in the studies listed in the table. Our choice of selected works is based on the use of modulation spectrogram related features, and use of advanced neural network architectures (e.g., contextual long short-term memory (LSTM), multi-head attention, ResNet architecture, multi-time-scale kernel, etc.). Below we briefly describe the details of selected works.

\begin{table}[t]
\caption{Performance comparison with selected works. \textbf{Boldface} values show the best obtained results.}

\centering
\hspace*{-1.75cm}
\begin{footnotesize}
\begin{tabular}{ccccccc}
\hline
\multirow{3}{*}{\textbf{References}} & \multirow{3}{*}{\begin{tabular}[c]{@{}c@{}}\textbf{Brief Description} \\ \textbf{(Feature \& Classifier)} \end{tabular}} &  \multicolumn{4}{c}{\begin{tabular}[c]{@{}c@{}}\textbf{Performance (in \%)}\end{tabular}} \\
 &  & \multicolumn{2}{c}{\begin{tabular}[c]{@{}c@{}}\textbf{EmoDB} \end{tabular}} & \multicolumn{2}{c}{\begin{tabular}[c]{@{}c@{}}\textbf{RAVDESS} \end{tabular}} \\
 &  & \textbf{Acc.} & \textbf{UAR} & \textbf{Acc.} & \textbf{UAR} \\  \hline \hline \\

\citeauthor{avila}~(\citeyear{avila}) & \begin{tabular}[c]{@{}c@{}}Different modulation \\ spectral measures. \\ DNN classifier.\end{tabular} & 55.40 & 46.83 & 37.77  & 29.10 \\ \\
\citeauthor{lightsernet}~(\citeyear{lightsernet}) & \begin{tabular}[c]{@{}c@{}} MFCC. \\ Parallel path fully \\ convolutional network. \end{tabular} & 73.37 & 67.21 & 48.68 & 44.51 \\ \\
\citeauthor{LIU20221}~(\citeyear{LIU20221}) & \begin{tabular}[c]{@{}c@{}} Interspeech 2009 feature set. \\ bc-LSTM~+~Multi-head attention. \end{tabular} & 53.45 & 47.49 & 24.58 & 21.79 \\ \\
\citeauthor{emonet}~(\citeyear{emonet}) & \begin{tabular}[c]{@{}c@{}} Mel-spectrogram. \\Convolutional ResNet. \end{tabular} & 76.34 & 71.07 & 52.56 & 49.46 \\ \\
\citeauthor{guizzo2020multi}~(\citeyear{guizzo2020multi}) & \begin{tabular}[c]{@{}c@{}} Spectrogram. \\ Multi-time-scale kernel \\ based CNN. \end{tabular}  & 63.01 & 58.03 & 40.27 &  36.46 \\ \\
~\citeauthor{PARRAGALLEGO2022103286}~(\citeyear{PARRAGALLEGO2022103286}) & \begin{tabular}[c]{@{}c@{}}x-vectors, i-vectors, INTERSPEECH \\ 2010 feature set, articulation, \\ phonation, and prosody features. \\ SVM classifier. \end{tabular} & 50.21 & 42.74 & 46.59 & 43.68 \\ \\
This work & \begin{tabular}[c]{@{}c@{}} GMT-MSF feature.\\ DNN-SVM framework. \end{tabular}  & 77.05 & 74.35 & \textbf{54.05} & \textbf{51.16} \\ \\
This work (Proposed method) & \begin{tabular}[c]{@{}c@{}} CQT-MSF feature. \\ DNN-SVM framework. \end{tabular} & \textbf{79.86} & \textbf{76.17} & 52.24 & 48.83 \\ \\ \hline

\end{tabular}
\label{comp_res_table}
\end{footnotesize}
\end{table}

\begin{itemize}

        \item Study performed by \citeauthor{avila}~(\citeyear{avila}) proposes feature pooling techniques over different measures of modulation spectral features for dimensional emotion recognition. For performance comparison, we compute the same modulation spectral measures over CQT representation, unlike \citeauthor{avila} which uses GMT representation, and show the results on our experimental framework. Feature-pooling schemes reported in the study are skipped to maintain similarity in comparison as our framework does not include handcrafted pooling operations.

        \item \citeauthor{lightsernet}~(\citeyear{lightsernet}) use mel-frequency cepstral coefficients (MFCCs) over a fully convolutional neural network architecture with parallel paths of different kernels sizes followed by stacked convolutional layers for SER with impressive reported SER performance. We use the GitHub implementation\footnote{\url{https://github.com/AryaAftab/LIGHT-SERNET}} of the state-of-the-art architecture with our experimental framework and CQT-MSF feature for performance comparison.

        \item We also select the state-of-the-art study performed by~\citeauthor{LIU20221}~(\citeyear{LIU20221}) for multi-modal emotion recognition in our work. As our primary focus is emotion recognition from speech, we use only the bidirectional-contextualised LSTM (bc-LSTM) with multi-head attention block used for speech modality in~\cite{LIU20221} with our experimental framework and databases. As emotion information spreads temporally across utterances, temporal pattern extraction architectures like LSTM, attention, are selected to compare with handcrafted temporal modulation feature, i.e., CQT-MSF.

        \item Study reported by~\citeauthor{emonet}~(\citeyear{emonet}) employs an adapter ResNet architecture for multi-corpora SER scenario. As our study does not include mixing different corporas, we select the reported ResNet architecture without the adapter module but with CQT-MSF feature for performance comparison. We use the open-source GitHub implementation\footnote{\url{https://github.com/EIHW/EmoNet}} of the model. 

        \item Study performed by~\citeauthor{guizzo2020multi}~(\citeyear{guizzo2020multi}) uses a multi-time-scale convolutional kernel based front-end which employs multiple temporally resampled versions of the original convolution kernel and parallelly perform convolution with every version. We utilize this method by employing multi-time-scale convolution layer as front-end over our DNN framework (mentioned in Section~\ref{classifier_desc_sec}). We use the available open-source GitHub implementation\footnote{\url{https://github.com/ericguizzo/multi_time_scale}} of multi-time-scale convolution layer. Similar to CQT-MSF, the multi-time-scale kernels also focus on the extraction of speech temporal patterns for emotion recognition.

\begin{figure}[t]
    \centering
    \includegraphics[scale=0.6]{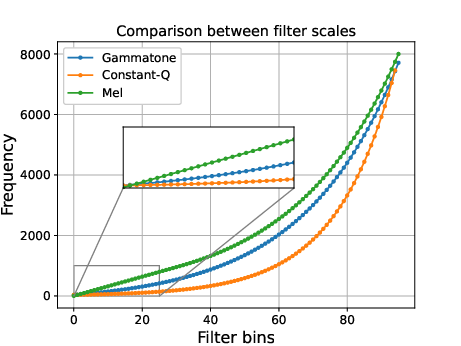}
    \caption{Comparison between different non-linear time-frequency representation scales with filterbank center frequencies (shown by dots). The zoomed-in portion shows the difference between the non-linearities for emotion-salient low frequency filter bins. For better visibility of differences among the scales, we plot every scale with $96$ frequency bins.}
    \label{filt_comp}
\end{figure}

        \item We select the feature set based study by \citeauthor{PARRAGALLEGO2022103286}~(\citeyear{PARRAGALLEGO2022103286}) for notably high performance reported on the RAVDESS database. The work includes combination of x-vector, i-vector, Interspeech $2010$ paralinguistic (IS10) feature set, and articulation, phonation and prosody features extracted from \emph{Disvoice}\footnote{\url{https://github.com/jcvasquezc/DisVoice}} framework for emotion classification with SVM as classifier. As the study report that the combination of only x-vector, IS10, and Disvoice framework provide the best performance, we use the same shortened feature set with SVM classifier, but on LOSO cross-validation framework. To maintain similarity, we use complete utterances (no segmentation) for both training and testing in this particular implementation. Note that study does not integrate our CQT-MSF and is rather based on the original implementation in the paper.

        \item As several modulation feature based SER works use gammatone-scale to generate a time-frequency representation over which modulations are computed~\cite{alam2013amplitude, avila, wu2011automatic}, we compare the performance of CQT-MSF with gammatone-scale based modulation spectral features (GMT-MSF) on the DNN-SVM framework mentioned in Section~\ref{classifier_desc_sec}. The obtained performance is reported in Table~\ref{comp_res_table} along with other employed comparison techniques. Our implementation of gammatone-spectrogram includes gammatone filter design using python~\emph{Spafe}~\cite{ayoub_malek_2022_6824667} toolkit, followed by application of the filters over the signal spectrogram. The GMT-MSF is computed over the designed gammatone time-frequency representation following the same steps used to compute CQT-MSF (Fig.~\ref{block_diagram}).

\end{itemize}

\begin{figure}[t]
    \centering
    \begin{minipage}{0.5\textwidth}
    \hbox{\hspace{-2cm}\includegraphics[scale=0.48]{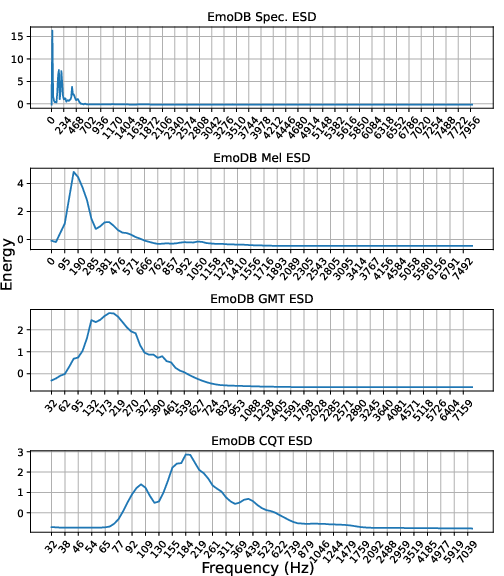}}
    \end{minipage}%
    \begin{minipage}{0.5\textwidth}
    \hbox{\hspace{0cm}\includegraphics[scale=0.48]{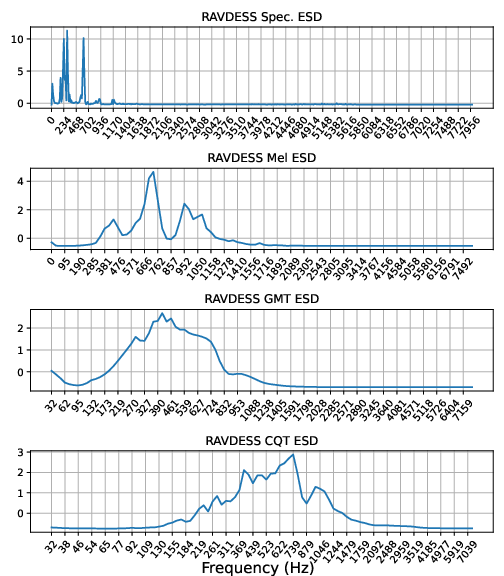}}
    \end{minipage}
    \caption{Average energy spectral density (ESD) computed over all utterances of EmoDB and RAVDESS. The spectrogram (Spec.), MFSC, CQT, and GMT ESD corresponds to the energy of corresponding coefficients average across all utterances. For better visibility of differences among the features, we plot every ESD with $96$ frequency bins.}
    \label{energy_curves}
\end{figure}

From Table~\ref{comp_res_table} we observe that CQT-MSF outperforms GMT-MSF on EmoDB database but performs relatively poor for RAVDESS database. This observation is due to the difference in non-linearity between constant-Q and gammatone scale. Previous studies on speech recognition~\cite{paliwalsfcc}, speaker recognition~\cite{SARANGI2020102795}, and SER~\cite{SINGH2022103712} also show how different non-linearity during frequency wrapping affect recognition performances. Inspired by those studies, we compare the three concerned non-linear scales used in our experiments: mel, constant-Q, and gammatone in Fig.~\ref{filt_comp} along with the filterbank center frequencies. For better visibility of differences in filter placement, we use $96$ frequency bins for every scale in this analysis. The figure shows that constant-Q scale provides highest low-frequency emphasis (because of binary logarithm) followed by gammatone and mel-scale. Considering the relevance of the low-frequency information for SER~\cite{singh2021, SINGH2022103712}, the underperformance of constant-Q scale is an interesting observation. \\

Previous studies indicate that the dataset-dependent non-linearity scale is more appropriate than a general-purpose scale and providing importance to higher energy region helps in performance optimization~\cite{paliwalsfcc, dipjyoti2017}. Hence, to further investigate the performance gap, we compare the energy spectral density (ESD) averaged across all utterances for both the databases. Fig.~\ref{energy_curves} shows the respective averaged energy density plots. The energy densities are computed by time averaging the squared feature coefficients, followed by averaging across all utterances. Fig.~\ref{energy_curves} shows that when compared to EmoDB, the energy density in RAVDESS is more shifted towards higher frequency regions. The constant-Q scale provides greater emphasis at low-frequency bins (refer Fig.~\ref{filt_comp}) but due to high non-linearity (binary logarithm), the resolution reduces drastically as we move towards higher frequencies. Hence, for RAVDESS, the resolution on the frequencies with the larger ESD value is lower in CQT compared to the resolution provided by gammatone filterbank placed in the ERB scale. Thus, for RAVDESS, GMT-MSF better captures the signal energy information and leads to better performance when compared to CQT-MSF. \\

From Table~\ref{comp_res_table} we also observe that the employed EmoNet-based ResNet architecture also shows competitive performance on RAVDESS database when compared to CQT-MSF feature. Larger model size with comparatively larger database leads to this observation.

\subsection{Visual Inspection of Learned Features} 
In spite of the performance gain achieved from the proposed CQT-MSF feature, the complexity of the model (Table~\ref{cnn_param_tab}) in terms of network parameters makes it very difficult to understand which information in input helps the network to recognize the pattern. To obtain a general insight into the operation of deep networks, several works use gradient generated at the ultimate layer with respect to the network input to generate a saliency map. This map shows corresponding input regions which weigh the most in generating the output probability scores~\cite{simonyan2014deep, springenberg2015striving, selvaraju2019gradcam}. We use one such analysis, called the \emph{gradient-based class activation mapping} (Grad-CAM), to obtain insight into the working of the model employed~\cite{selvaraju2019gradcam}. \\

\renewcommand{\thesubfigure}{\alph{subfigure}}
\begin{figure}[hp!]

    \begin{subfigure}[t]{\textwidth}
    \hbox{\hspace{1cm}\includegraphics[height=6cm, width=11cm]{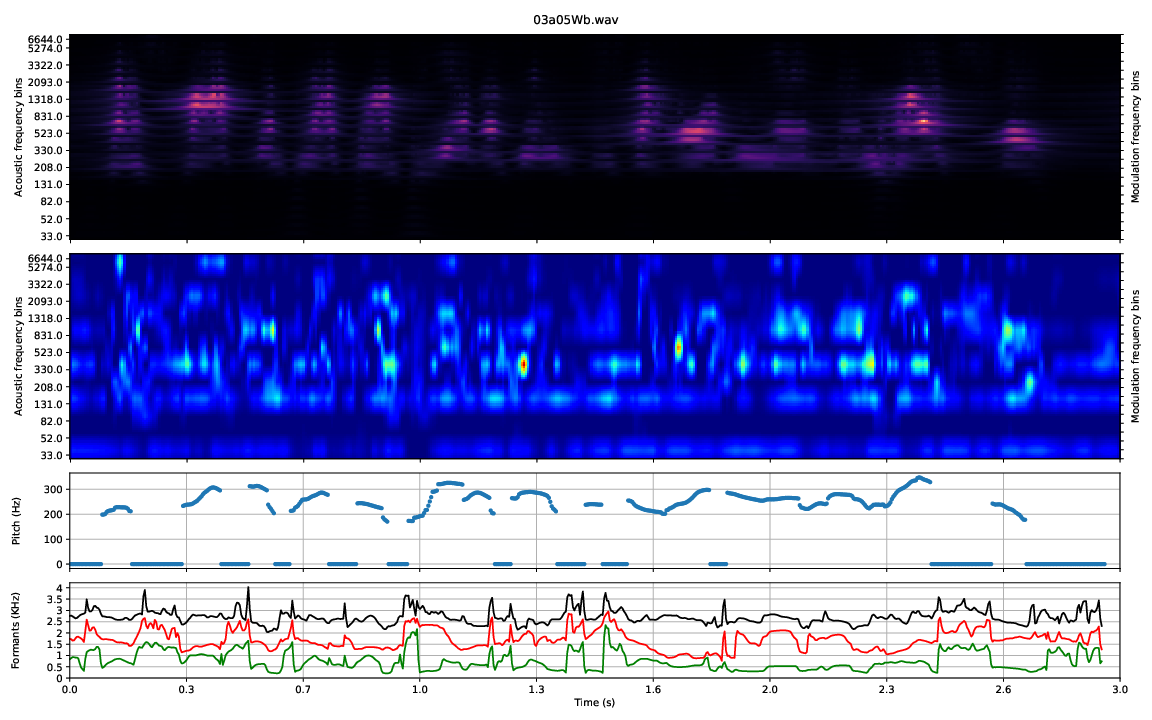}}
    \caption{Constant-Q MSF with Grad-CAM, pitch and first three formants of \emph{Anger} utterance}
    \end{subfigure}
    
    \begin{subfigure}[t]{\textwidth}
    \hbox{\hspace{1cm}\includegraphics[height=6cm, width=11cm]{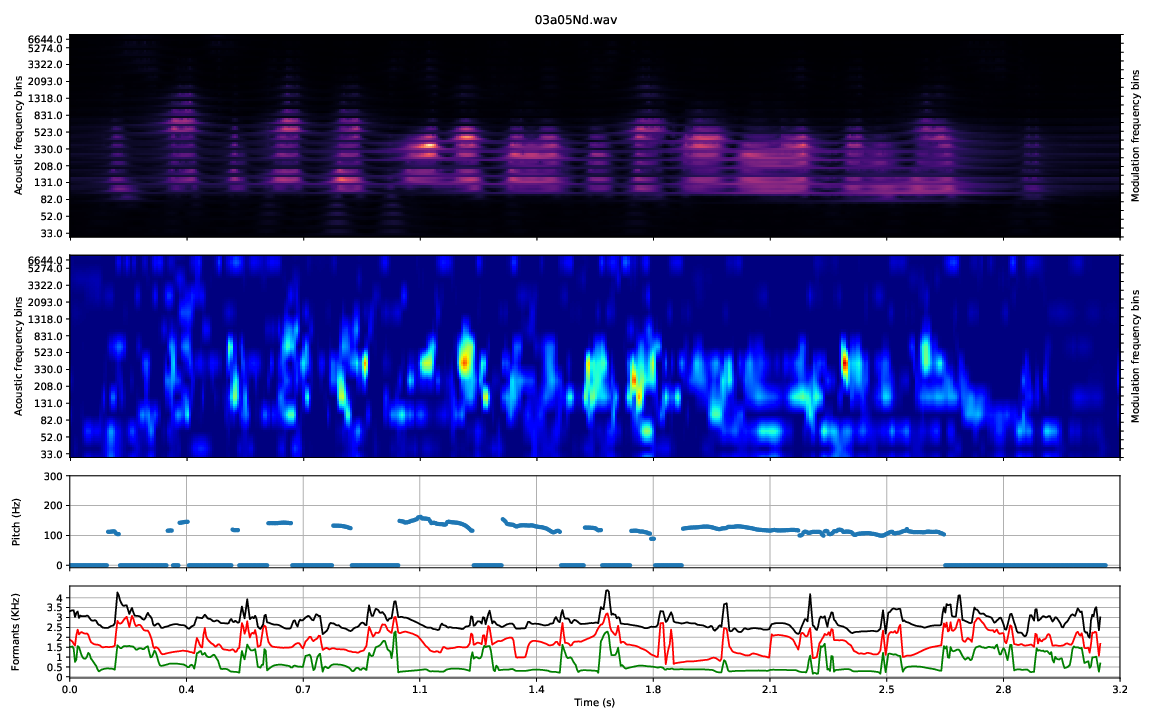}}
    \caption{Constant-Q MSF with Grad-CAM, pitch and first three formants of \emph{Neutral} utterance}
    \end{subfigure}
    
    \begin{subfigure}[t]{\textwidth}
    \hbox{\hspace{1cm}\includegraphics[height=6cm, width=11cm]{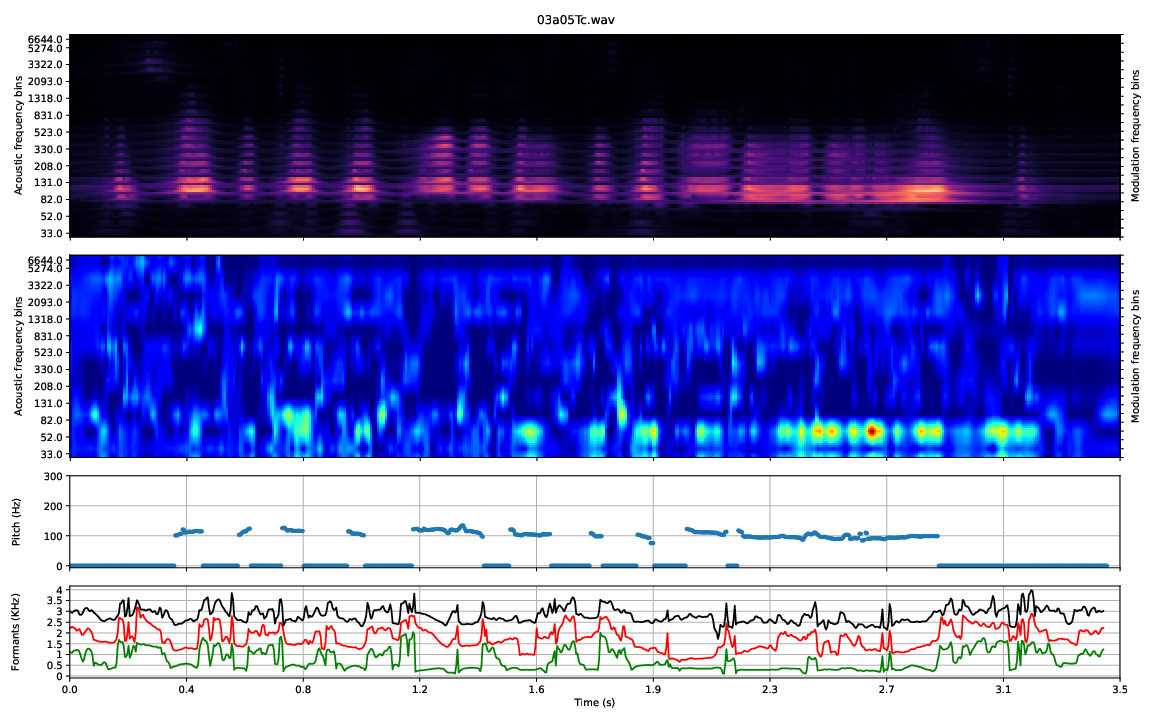}}
    \caption{Constant-Q MSF with Grad-CAM, pitch and first three formants of \emph{Sad} utterance}
    \label{pitch_gradcam}
    \end{subfigure}
    
    \caption{Grad-CAM output of three different emotion utterances of EmoDB database. To analyse the significance of various frequency regions, pitch and first three formants are also shown along with the Grad-CAM output. For MSF plot, the y-axis labels on the left describe the auditory frequencies and ticks on the right describe the modulation bins corresponding to every auditory frequency bin. The title of every plot describes the EmoDB file name used for analysis.}
    \label{gradcam}
\end{figure}

Grad-CAM uses the class-wise gradient generated at network output with respect to the activations of the final convolution layer to generate a \emph{heatmap} showing the importance of different regions of the input. The steps included in Grad-CAM heatmap generation are:
\begin{enumerate}
    
    \item Compute the gradient of network output score (before softmax non-linearity) with respect to the activations of final convolution layer, i.e., $\frac{\partial y_c}{\partial A_k}$, where $y_c$ is the score of the $c$th class and $A_k$ is the 3-dimensional (height, width and channel dimension) class activations from the final convolution layer.
    
    \item Average the computed gradients over length and width dimensions (global average pooling) to obtain a single vector representation of gradients. Mathematically, $\alpha_k~=~\frac{1}{N} \sum \limits_{\mathrm{length}} \sum \limits_{\mathrm{width}} \frac{\partial y_c}{\partial A_k}$.
    
    \item Multiply the computed gradient vector with the final convolution layer activations and average the result over the number of filters in convolution layer, i.e., $map~=~\frac{1}{N} \sum \limits_k(\alpha_k A_k) $.
    
    \item Apply \emph{ReLU} activation over computed class activation maps in the previous step, $L^{c}_{\mathrm{Grad}-\mathrm{CAM}}~=~ReLU(map)$.
    
    \item Upsample the computed $2$-D heatmap over length and width axes to make its shape similar to the input image. The upsampled heatmap shows the importance of various regions of input image which led to the final class prediction.

\end{enumerate}

Fig.~\ref{gradcam} shows the Grad-CAM output of the CQT-MSF feature of \emph{Anger}, \emph{Neutral}, and \emph{Sad} utterances of the same speaker (speaker $03$) and context (a$05$) from EmoDB database. For analysis of Grad-CAM output, we also plot the pitch and first three formant frequencies (F$1$, F$2$ \& F$3$) of utterances to compare the frequency regions which are most focused upon by the network to predict emotion classes. We observe that pitch and the first two formants (F$1$ \& F$2$) are important for the emotion class prediction. Lower pitch harmonics are apparently more significant for \emph{Sad} emotion as shown in Fig.~\ref{pitch_gradcam}. Another important observation for \emph{Sad} is the presence of high Grad-CAM score at the silence and unvoiced regions (where pitch frequency is $0$~Hz) of utterance. This shows that the silence between spoken phonemes is also important for emotion recognition. In \emph{Anger} and \emph{Neutral} emotions, formants F$1$ \& F$2$ are more prominent than pitch. Both emotions are identified mostly in the voiced region of utterances. Interestingly, the Grad-CAM response of \emph{Sad} also shows some focus on high frequencies (near formant F3) over complete utterance as compared to \emph{Neutral} emotion class. Observations made from the Grad-CAM analysis indicate the importance of low frequencies for emotion recognition, especially for low-arousal emotions like \emph{Sad}, further justifying the use of CQT time-frequency representation for SER. Also, MSF representation provides the Conv2D classifier with different modulation rates of different speech characteristics, such as pitch harmonics, formants etc. This helps the classifier to emphasize the emotion-wise differences appearing across different modulation rates, for different speech characteristics (pitch, formant, etc.), hence improving the performance.

\subsection{Discussion}
Our performed experiments show higher relevance of CQT-based features, as compared to mel-scale features, for SER. We summarise and interpret the results obtained from the experiments in the following points:

\begin{itemize}
    \item \textbf{The two-staged auditory processing based features, i.e., early auditory with cortical analysis based features improve emotion classification performance}. This also justifies the combination of human auditory analysis (domain knowledge) with neural networks for the betterment of SER. However, such improvement is observed over temporal modulations extracted from CQT representation only. Temporal modulations of MFSC show opposite effect and degrade the performance. 
    
    \item \textbf{The improvement observed with CQT-MSF (both fusion types) and standalone CQT over MFSC features show the relevance of increased low-frequency resolution in CQT for SER}. As low-frequency resolution in mel-scale is not as high, it does not provide enough emphasis over the emotion relevant low-frequencies to capture the information required for emotion discrimination. Hence, its modulation spectrum coefficients also end up with less emotion relevant parts of speech, e.g., irrelevant high frequency regions, leading to reduction in performance.
    
    \item \textbf{DNN framework lags behind in performance as compared to DNN-SVM classification framework with RBF kernel function}. This observation is consistent across every employed feature. The advantage in performance of DNN-SVM framework has also been mentioned in SER \cite{zhang2017speech} and speaker recognition \cite{snyder2018x} works.
    
    \item \textbf{The confusion matrices of different features show a general misclassification trend in \emph{Happy-Anger} and \emph{Fear-Happy} emotion pairs}. This confusion mainly appears due to very similar arousal characteristics of features. However, \emph{Happy-Anger} pair are placed very distant in valence plane. This is due to higher focus on speech prosody in constant-Q representation, and higher sensitivity of speech prosody over arousal characteristics~\cite{iker2010}. Although, inclusion of modulations of constant-Q representation reduces the confusion among emotions with opposite valence characteristics.
    
    \item \textbf{Both CQT-MSF and standalone CQT also outperforms scattering transform coefficients}. Scattering transforms apply averaging over features with empirically defined averaging scale to obtain a translation invariant representation. The convolutional neural network, when used with constant-Q features as input, automatically learns this requires invariance with cross-entropy objective function. Hence, the joint effect of CQT/CQT-MSF and automatic translation invariance makes our frameworks superior. Although, scattering transform manages to outperform mel-scale features, again because of the constant-Q filter banks and time shift and deformation stability in scattering coefficients.

    \item \textbf{CQT-MSF feature outperforms GMT-MSF on EmoDB but underperforms on RAVDESS database.} Comparative analysis shows a slight high-frequency shift in average energy spectral density of the RAVDESS database compared to EmoDB database. The difference in non-linearities of the gammatone and CQT scales result in gammatone-spectrogram better capturing energies with a slight high-frequency shift. This explains the observed anomaly in the performance with RAVDESS database.

\end{itemize}

Studies in psychology report that individuals with music expertise are better capable of perceiving emotions from speech \cite{lima2011speaking, good2017benefits, fuller2014musician, dmitrieva2006ontogenetic, twaite2016examining, weijkamp2017attention, thompson2004decoding, nussbaum2021}. This finding falls in line with our experiments. As CQT was originally invented for music analysis, its better suitability for SER can be considered as a mathematical evidence, supporting the findings in psychology domain. Another justification towards increased SER suitability of CQT, can be proposed by analysing studies performed over \emph{amusia} in \cite{sydney2021, nussbaum2021}. Amusia is a medical condition in which individuals have limited capability to perceive or resolve pitch. The study in \cite{sydney2021} reports that emotion recognition ability of amusic individuals is below par with that of normal individuals, which is attributed to their limited ability to resolve pitch or low-frequencies of speech. Amusics then utilize the high-frequency content of speech to decipher emotions but are not as efficient as healthy individuals. Therefore, an amusic brain can be assumed to represent speech as a time-frequency representation with low resolution at low-frequencies and comparatively higher resolution at high frequencies. In a contrastive manner, a representation with high low-frequency resolution should improve SER ability, which is what we observe in our experiments with CQT-based features. Also, modulation coefficients computed over CQT is analogous to cortex-level analysis performed over music trained brain. Study performed in~\cite{nussbaum2021} reports that such cortical analysis has further beneficial effects over SER ability.

\section{Conclusion}
\label{conc_sec}

This paper proposed the use of constant-Q transform based modulation spectral features for SER. The proposed feature employs the knowledge of two-staged sound processing in humans as domain knowledge and is tested over two different deep network based classification frameworks. We show that the proposed feature outperforms standard mel-frequency based feature and scattering transform coefficients. From the performed experiments, we conclude the following:

\begin{enumerate}

    \item A representation with increased low-frequency resolution is a better contender for SER. Similar conclusion is endorsed in psychology based studies as well.

    \item The combination of a time-frequency representation with higher low-frequency resolution, and its temporal modulation (two-staged representation), efficiently represent the emotion contents in speech.

    \item Mel-scale based feature and its temporal modulations are not very significant from speech emotion information perspective.

    \item Similar, to mel-scale features, CQT and its temporal modulation representation are also time deformation invariant. With CQT-MSF as input, the CNN can learn the required invariance to time-frequency shifts, leading to a representation stable to shifts and deformations.

    \item The DNN-SVM framework provides better SER performance as compared to the standard DNN framework. 

    \item Grad-CAM based analysis performed over MSF reestablishes the importance of pitch and formant frequencies for SER. It also describes the importance of different modulation rates of pitch and formants, apart from their crude values, for SER.

\end{enumerate}    

Although the proposed feature performs better than standard features, the performance is still not optimum for practical real-world deployment of the feature. Also, the performance varies by a large margin when the feature is used over different databases. Even though found more efficient than mel-scale, constant-Q scale is effective for utterances with larger average energy in low-frequency regions. This opens up the opportunity to explore database-dependent non-linear scale for SER. In future, we would also like to experiment with joint spectral and temporal modulation feature and analyse its suitability for SER. To combine the domain knowledge with self-learning, a deep network based architecture for self-learned modulation feature extraction can also be explored.

\bibliography{mybib}

\end{document}